\documentclass[floats,floatfix,showpacs,amssymb,prd,twocolumn,superscriptaddress,nofootinbib,nolongbibliography,reprint,preprintnumbers]{revtex4-1}

\usepackage{amssymb,amsmath,verbatim,mathtools,needspace,enumitem,etoolbox,graphicx,physics,microtype,afterpage,xspace,tabularx,lmodern,multirow,braket}
\usepackage[justification=raggedright,singlelinecheck=false]{caption}
\usepackage{gensymb}
\usepackage[normalem]{ulem}
\usepackage[dvipsnames, usenames]{xcolor}
\definecolor{linkcolor}{rgb}{0.0,0.3,0.5}
\usepackage[unicode, colorlinks=true, linkcolor=linkcolor, citecolor=linkcolor, filecolor=linkcolor, urlcolor=linkcolor, linktocpage, breaklinks]{hyperref}
\usepackage[all]{hypcap}
\usepackage{subfigure}
\usepackage[T1]{fontenc}
\usepackage[utf8]{inputenc}
\usepackage[usenames,dvipsnames]{xcolor}
\usepackage{booktabs}
\hypersetup{colorlinks=true,citecolor=romared,linkcolor=romared,urlcolor=romared}

\definecolor{romared}{RGB}{142,0,28}

\newcommand{\be}{\begin{equation}}
\newcommand{\ee}{\end{equation}}

\def\be{\begin{equation}}
\def\ee{\end{equation}}
\newcommand{\beq}{\begin{eqnarray}}
\newcommand{\eeq}{\end{eqnarray}}

\usepackage{aas_macros}
\usepackage{makecell}
\usepackage{soul}
\usepackage{amssymb}

\usepackage{lipsum}
\usepackage{physics}
\usepackage{graphicx}
\usepackage{caption}
\usepackage{float}

\newcolumntype{Y}{>{\centering\arraybackslash}X}

\begin{document}

\title{Exceptional Lines and Excitation of (Nearly) Double-Pole Quasinormal Modes:\\A Semi-Analytic Study in the Nariai Black Hole}

\author{Nao Nakamoto} 
\affiliation{Department of Physics, Kyoto University, Kyoto 606-8502, Japan}
\author{Naritaka Oshita}
\affiliation{Center for Gravitational Physics and Quantum Information, Yukawa Institute for Theoretical Physics, Kyoto University, 606-8502, Kyoto, Japan}
\affiliation{The Hakubi Center for Advanced Research, Kyoto University,
Yoshida Ushinomiyacho, Sakyo-ku, Kyoto 606-8501, Japan}
\affiliation{RIKEN iTHEMS, Wako, Saitama, 351-0198, Japan}
\affiliation{Department of Physics, Kindai University, Osaka 577-8502, Japan}

\preprint{YITP-25-199, RIKEN-iTHEMS-Report-26}

\begin{abstract}
We show that quasinormal modes (QNMs) of a massive scalar field in Kerr-de Sitter and Myers-Perry black holes exhibit an exceptional line (EL), which is a continuous set of exceptional points (EPs) in parameter space, at which two QNM frequencies and their associated solutions coincide.
We find that the EL appears in the parameter space spanned by the scalar mass and the black hole spin parameter, and also in the Nariai limit, i.e., $r_{\rm c} - r_{\rm h} \to 0$, where $r_{\rm c}$ and $r_{\rm h}$ denote the radii of the cosmological and black hole horizons, respectively.
We analytically study the amplitudes or excitation factors of QNMs near the EL.
Such an analytic treatment becomes possible since, in the Nariai limit, the perturbation equation reduces to a wave equation with the P\"{o}schl-Teller (PT) potential.
We discuss the destructive excitation of QNMs and the stability of the ringdown near and at the EL.
The transient linear growth of QNMs---a characteristic excitation pattern near an EP or EL---together with the conditions under which this linear growth dominates the early ringdown, is also studied analytically.
Our conditions apply to a broad class of systems that involve the excitation of (nearly) double-pole QNMs.

\end{abstract}

\maketitle

%%%%%%%%%%%%%%%%%%%%%%%%
\section{Introduction}
%%%%%%%%%%%%%%%%%%%%%%%%

Gravitational waves from the merger of binary black holes (BHs) enable us to test general relativity in the strong-field regime.
In particular, the ringdown phase---the relaxation process through which the remnant BH settles into a stationary state---is characterized by a set of discretized spectra of damped oscillations known as quasinormal modes (QNMs) \cite{Kokkotas:1999bd, Berti:2009kk, Berti:2025CQG}.
The frequencies and damping rates of QNMs depend only on the intrinsic properties of the BH---its mass, spin, and charge.
Precise measurements of these modes through BH spectroscopy can verify the no-hair theorem and potentially uncover deviations from general relativity or signatures of new physics.

Recently, QNM properties analogous to those of non-Hermitian systems have been actively studied, particularly in the context of spectral instability or pseudospectrum \cite{Nollert:1996rf,Barausse:2014tra,Jaramillo:2020tuu,Jaramillo:2021tmt,Cheung:2021bol,Berti:2022xfj}, as well as avoided crossings (ACs) and exceptional points (EPs) \cite{Dias:2021yju,Davey:2022vyx,Dias:2022oqm,Davey:2023fin,Motohashi:2024fwt,Cavalcante:2024swt,Oshita:2025ibu,Lo:2025njp,Yang:2025dbn,Kubota:2025hjk,Cao:2025afs,Wu:2025wbp}.
The pseudospectrum is related to the high sensitivity of the QNM spectrum to the environmental effects or small corrections in the perturbation equation.
The AC refers to the phenomenon in which two QNM branches approach each other in frequency but repel as the BH parameters vary.
With sufficient fine-tuning of the parameters, the two modes degenerate and form a {\it double-pole QNM}.
A parameter set that leads to such a degeneracy in QNM frequencies is referred to as an EP.
The excitation factors of QNMs \cite{Leaver:1986gd,Sun:1988tz,Andersson:1995zk,Glampedakis:2001js,Glampedakis:2003dn,Berti:2006wq,Zhang:2013ksa,Oshita:2021iyn} (see also the recent works Refs.~\cite{Oshita:2024wgt, Motohashi:2024fwt,Lo:2025njp,Kubota:2025hjk,DellaRocca:2025zbe}), which are the residues of the Green's function at QNM poles and quantify the excitability of QNMs, are enhanced by AC \cite{Motohashi:2024fwt} and diverge at the degeneracy of two modes.
Nevertheless, time-domain ringdown waveforms are insensitive or stable with respect to pseudospectral effects \cite{Nollert:1996rf,Barausse:2014tra,Jaramillo:2020tuu,Jaramillo:2021tmt,Berti:2022xfj} and to ACs or EPs \cite{Oshita:2025ibu}.
As analytically shown in Ref.~\cite{Oshita:2025ibu}, the huge enhancement of excitation factors due to AC or EP interferes destructively, and the resulting superposed waveform remains stable.
\begin{figure}[t]
\centering
\includegraphics[width=0.45\textwidth]{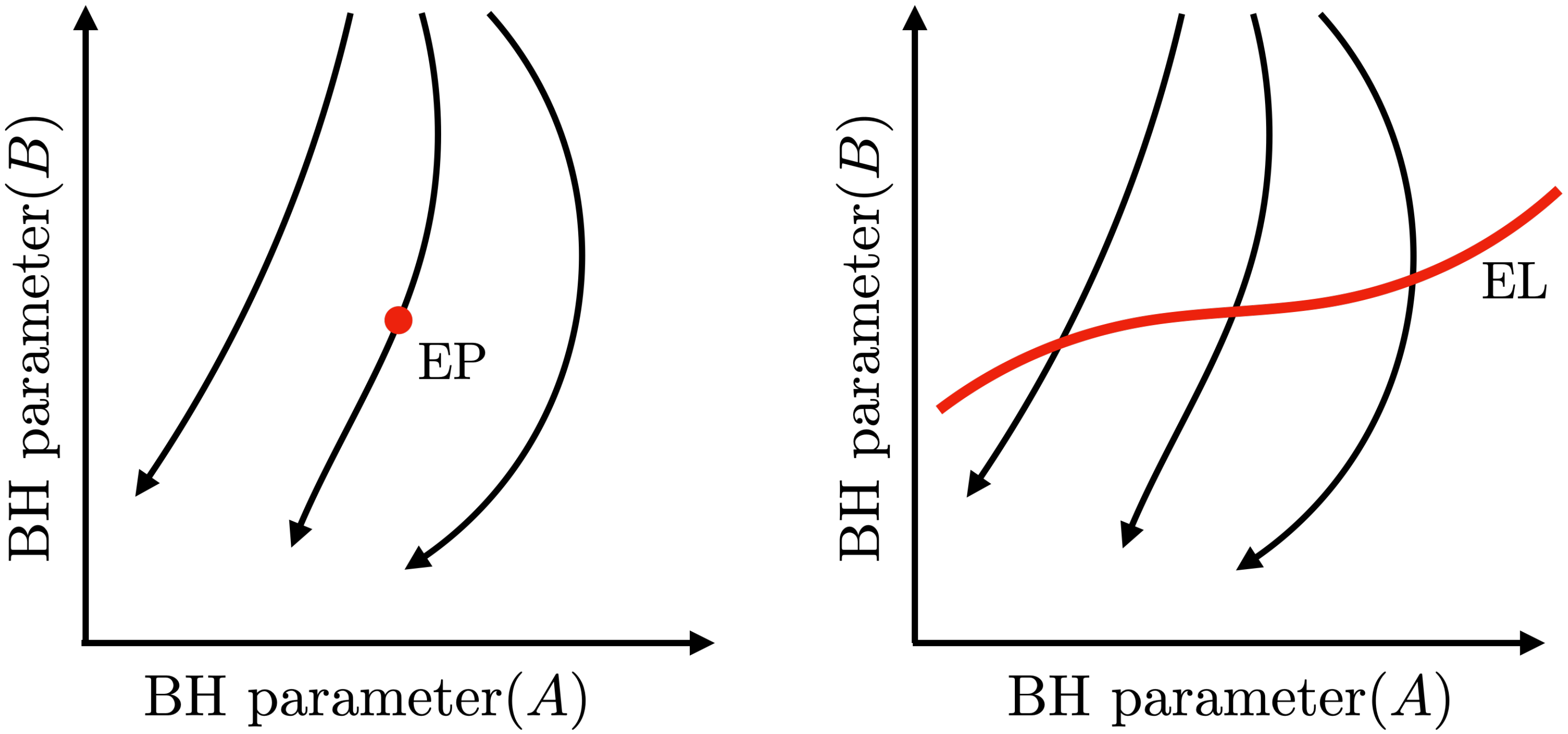}
\caption{
Schematic picture of BH parameter trajectories (black arrows) in a relevant parameter space spanned by, say, $A$ and $B$, involving an EP (left) and an EL (right). The example assumes that the relevant parameter space is restricted to two dimensions.
}
\label{fig_EPEL_schematic}
\end{figure}
A possible unique feature of QNM excitation, i.e., transient linear growth at early ringdown, associated with AC was identified in Ref.~\cite{Yang:2025dbn}.
In Ref.~\cite{Yang:2025dbn}, it was pointed out that the linear growth of the superposed QNM amplitude arises from the beating between the two modes involved in the AC, which possess slightly different QNM frequencies.
Although this linear growth is eventually suppressed by the exponential damping of QNMs and may be contaminated by the excitation of other single-pole modes, it remains an interesting phenomenon.
A fitting analysis specifically designed to capture this feature was also recently developed in Ref.~\cite{PanossoMacedo:2025xnf}.
While most studies so far have focused on EPs, realizing an EP typically requires delicate fine-tuning of the system parameters, which limits its physical relevance.
In contrast, if an exceptional line (EL) forms a one-dimensional locus in the relevant parameter space, it allows the degeneracy of two QNMs---and hence a double-pole QNM excitation---to occur with reduced fine-tuning of the BH parameters, depending on the physical setup (see FIG.~\ref{fig_EPEL_schematic}).
Recently, an EL was indeed discovered by introducing a bump correction in the Regge-Wheeler equation in Ref.~\cite{Cao:2025afs}.
Of course, when many BH parameters are involved, the relevant parameter space can still be considerably large, and severe fine-tuning may remain unavoidable, even in the presence of an EL.
Therefore, the advantage of having an EL becomes more significant when the relevant parameter space is effectively low-dimensional.

In this work, we propose that the perturbation of a massive scalar field around a single rotating Myers-Perry-de Sitter (MP-dS) BH in the Nariai limit \cite{Nariai1950} provides a (semi-)analytically tractable model for ELs of QNMs.
We note that the Nariai BH with a cosmological constant is an idealized configuration and should not be regarded as a realistic astrophysical system.
Nevertheless, it offers a clean theoretical laboratory in which the excitation and stability properties of (nearly) double-pole QNMs and ELs can be investigated (semi-)analytically.

The computation of QNM frequencies typically requires numerical analysis, since perturbation equations rarely admit closed-form solutions.
For this reason, a variety of simplified setups, e.g., the P\"{o}schl-Teller (PT) potential \cite{Poschl1933}, have been employed as toy models to gain analytic insight into their qualitative features \cite{Berti:2009kk, Shu:2004fj}.
Our work is a natural extension of Ref.~\cite{Cardoso:2003sw}, which showed that a massless field in the Schwarzschild-de Sitter spacetime in the Nariai limit is governed by the plane-wave scattering at a PT potential barrier, allowing QNM frequencies to be obtained in closed form.
By introducing a massive scalar field and considering the spin of a $d$-dimensional BH---both of which still allow closed-form QNM expressions in the Nariai limit \cite{Ponglertsakul:2020ufm}---we analytically investigate the degeneracy of QNMs at EP, which leads to a double-pole QNM, and study its excitation properties.

This paper is organized as follows.
In Sec.~\ref{Sec_massive_scalar_MP}, we introduce the perturbation equations of a massive scalar field in an MP-dS spacetime.
In Sec.~\ref{Sec_mode_excitation}, we show that the radial perturbation equation reduces to a PT-type equation parametrized by the scalar mass $\mu$ and the surface gravity $\kappa_{\rm h}$.
We analytically demonstrate that the solution to the PT-type equation exhibits an EP, and the double-pole QNM shows linear growth in the time domain, which is eventually suppressed by the exponential damping.
We then explore EPs for four-dimensional non-spinning and spinning BHs in the Nariai limit.
It turns out that this system exhibits an EL in the parameter space of the scalar mass $\mu$ and black-hole spin parameter $a$.
Sec.~\ref{sec_conclusions} is devoted to conclusions.
In Appendix~\ref{App_d5_case}, we present a similar analysis for a higher-dimensional ($d = 5$) BH. We confirm that the five-dimensional system also admits an EL.
In Appendix~\ref{Sec_div_EFs}, we also discuss the divergence of QNM excitation factors at EPs in an analytic way.

%%%%%%%%%%%%%%%%%%%%%%%%
\section{Perturbation of a massive scalar field in the MP-dS spacetime}
\label{Sec_massive_scalar_MP}
%%%%%%%%%%%%%%%%%%%%%%%%

%%%%%%%%%%%%%%%%%%%%%%%%
\subsection{MP-dS spacetime}
%%%%%%%%%%%%%%%%%%%%%%%%

We here investigate the scalar perturbation of a $d$-dimensional MP-dS BH with a single rotation of a spin parameter $a$ and a cosmological constant $\Lambda$. 
The metric of the MP-dS BH can be written as
\begin{align}
\begin{split}
    ds^2 &= g_{\mu \nu} dx^{\mu} dx^{\nu} \\
    &= -\frac{\Delta_r}{\rho^2\Sigma^2}\qty(\Delta_\theta dt - a\sin^2\theta d\Phi)^2 + \frac{\rho^2}{\Delta_r}dr^2 + \frac{\rho^2}{\Delta_\theta}d\theta^2 \\
    &+ \frac{\Delta_\theta\sin^2\theta}{\rho^2\Sigma^2}\qty(a(1-\Lambda r^2)dt-(r^2+a^2)d\Phi)^2 \\
    &+ r^2\cos^2\theta d\Omega^2_{d-4},
\end{split}
\end{align}
where
\begin{align*}
    \rho^2 &\coloneqq r^2 + a^2\cos^2\theta, \\
    \Delta_r &\coloneqq (r^2+a^2)(1-\Lambda r^2)-2Mr^{5-d}, \\
    \Sigma &\coloneqq 1+\Lambda a^2, \\
    \Delta_\theta &\coloneqq 1+\Lambda a^2\cos^2\theta, \\
    d\Omega^2_{d-4} &\coloneqq \sum_{i=1}^{d-4}\qty(\prod_{j=1}^{i}\sin^2\varphi_{j-1})d\varphi^2_i, \quad\varphi_0 \coloneqq \frac{\pi}{2}.
\end{align*}

\begin{figure*}[t]
\centering
\includegraphics[width=1\textwidth]{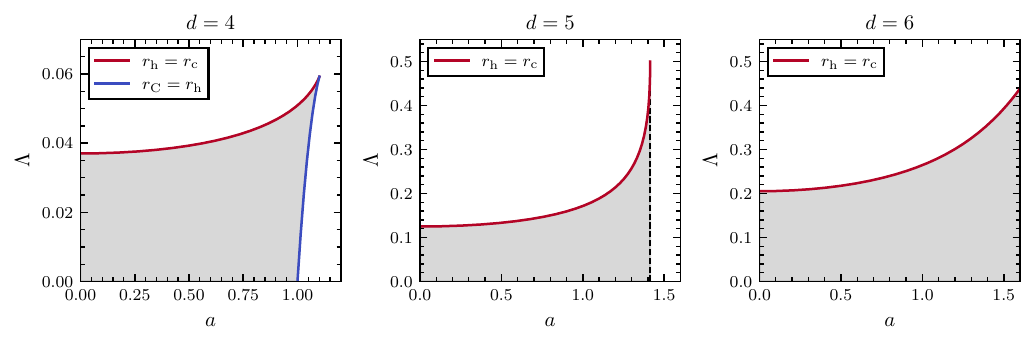}
\caption{
Phase space of the MP-dS BH with a single rotation and $M=1$. The spacetime dimension is set to $d=4$ (left), $d=5$ (center), and $d=6$ (right). 
The black dashed line stands for the upper bound in \eqref{upper_spin_d5}.
The BH solution exists in the gray-shaded region.
}
\label{PS}
\end{figure*}
The spacetime possesses the inner, outer, and cosmological horizons at $r = r_{\rm C}$, $r_{\rm h}$ and $r_{\rm c}$, respectively, with $r_{\rm C} \leq r_{\rm h} \leq r_{\rm c}$, corresponding to the roots of $\Delta_r=0$.
In this setting, the horizon velocity and surface gravity at each horizon are
\begin{align}
    \Omega_i &= \frac{a(1-r_i^2\Lambda)}{r_i^2+a^2}, \\
    \kappa_i &= \frac{1}{2(r_i^2+a^2)}\frac{d\Delta_r(r)}{dr}\bigg|_{r=r_i}\,,
\end{align}
where the subscript $i = \text{C, h, c}$ indicates the quantities associated with the inner horizon, BH horizon, and the cosmological horizon, respectively.

The three parameters, BH mass $M$, spin parameter $a$, and cosmological constant $\Lambda$, characterize the BH.
Since this spacetime has three horizons for $d\ge4$ in general, the following two extremal configurations may arise: $r_{\text{C}}=r_{\text{h}}$ or $r_{\text{h}}=r_{\text{c}}$. 
The parameter conditions for the extremal limits depend on the dimension of the spacetime.
As a reference, the phase diagrams of the parameter conditions for $d = 4$, $5$, and $6$ are shown in FIG. \ref{PS}.
The case of $d=4$ admits the two types of extremal configurations, i.e., $r_{\rm C} = r_{\rm h}$ (blue line in FIG.~\ref{PS}) and $r_{\rm h} = r_{\rm c}$ (red line in FIG.~\ref{PS}). 
For dimensions $d = 5$ and $d = 6$, only  $r_{\text{h}}=r_{\text{c}}$ is possible. In the case of $d=5$, there exist BH solutions for the spin parameter satisfying \cite{Kodama:2009rq}
\begin{align}
    0\le |a| \le \sqrt{2M}.
\label{upper_spin_d5}
\end{align}
In the case of $d\ge6$, there is no restriction on the spin parameter. Throughout this paper, we set $M=1$.

%%%%%%%%%%%%%%%%%%%%%%%%
\subsection{Scalar perturbation}
%%%%%%%%%%%%%%%%%%%%%%%%

Let us consider the perturbation of a scalar field with mass $\mu$ governed by the Klein-Gordon (KG) equation:
\begin{align}\label{fieldeq}
    \frac{1}{\sqrt{-g}}\partial_\mu \qty(\sqrt{-g}g^{\mu\nu}\partial_\nu\Psi) - \mu^2\Psi = {\cal T}\,,
\end{align}
where ${\cal T}$ is the source term.
We introduce the separation of variable ansatz:
\begin{align}
    \Psi =  \int_{- \infty}^{\infty} \frac{\dd \omega}{2 \pi} e^{-i\omega t}e^{im\Phi}R_{kjm}(r)S_{kjm}(\theta)Y_{jd}(\varphi_1,\cdots,\varphi_{d-4})\,,
\end{align}
where $m$ is the azimuthal parameter and $\omega$ is frequency conjugate to $t$. 
Substituting the ansatz into the original equation (\ref{fieldeq}) yields three separated equations for $Y_{jd}$, $R_{kjm}$, and $S_{kjm}$. The equation with respect to $\varphi_i$ is the hyperspherical harmonics about $(d-4)$-dimensional Laplacian \cite{Boonserm:2014fja}:
\begin{align}
    \Delta_S Y_{jd} &\equiv \sum_{i=1}^{d-4}\frac{g^{\varphi_i\varphi_i}}{(\sin\varphi_i)^{d-4-i}}\partial_{\varphi_i}((\sin\varphi_i)^{d-4-i}\partial_{\varphi_i}Y_{jd}) \notag\\
    &= -j(j+d-5)Y_{jd}. \label{LaplaceY}
\end{align}
For $d=4$, we can take $Y_{jd}=1$.
The KG equation can be separated into the radial and angular parts by introducing a separation constant. 
The radial and angular equations are
\begin{align}\label{Req1}
    &\frac{1}{r^{d-4}}\partial_r[r^{d-4}\Delta_r\partial_r R_{kjm}] \notag \\ &+ \bigg[\frac{1}{\Delta_r}(\omega(r^2+a^2)-ma(1-\Lambda r^2))^2 -\mu^2r^2 \notag\\
    & -\frac{j(j+d-5)a^2}{r^2}-A_{kjm} \bigg]R_{kjm}= T_{kjm}\,,
\end{align} 
\begin{align}\label{Seq1}
    &\frac{1}{\sin\theta\cos^{d-4}\theta}\partial_\theta[\sin\theta\cos^{d-4}\theta\Delta_\theta\partial_\theta S_{kjm}] \notag\\
    &+ \bigg[-\frac{1}{\Delta_\theta}\bigg(a\omega\sin\theta-\frac{m\Delta_\theta}{\sin\theta}\bigg)^2
    -\mu^2a^2\cos^2\theta \notag\\&
    -\frac{j(j+d-5)}{\cos^2\theta} + A_{kjm}\bigg]S_{kjm}=0\,,
\end{align}
where $T_{kjm}$ is the source term in the frequency space, and $A_{kjm}$ is the eigenvalue of the separation constant for which the spheroidal harmonics $S_{kjm}(\theta)$ is regular at the poles.
The eigenvalue has the integer indices $k$, $j$, and $m$, whose details will be described later. In general, the eigenvalue depends on $d, a, \omega, \Lambda$, and $\mu$ \cite{Ponglertsakul:2020ufm, Cho:2009wf, Berti:2005gp}. 
In the following, we will consider methods for solving the two equations in the Nariai limit.
Unless otherwise stated, we omit the subscripts $k,j,m$ of variables for simplicity.

%%%%%%%%%%%%%%%%%%%%%%%%
\subsection{Radial Equation in the Nariai limit}
%%%%%%%%%%%%%%%%%%%%%%%%

Let us introduce a new radial perturbation variable $\bar{R}$:
\begin{align}
    \bar{R} \coloneqq R \sqrt{r^{d-4}(r^2+a^2)}.
\end{align}
The radial Eq. (\ref{Req1}) becomes
\begin{align}
    &\frac{\dd^2{\bar{R}}}{\dd{r^*}^2} + \bigg\{ (\omega-m\Omega)^2 + \frac{\Delta_r}{(r^2+a^2)^2}\bigg[ -A_{kjm}-\mu^2r^2 \notag\\
    &-\frac{j(j+d-5)a^2}{r^2}+\Delta_r \bigg(\frac{2r^2-a^2}{(r^2+a^2)^2}-\frac{(d-4)(d-6)}{4r^2} \bigg) \notag \\
    &-\frac{\Delta^{\prime}_r}{2r}\bigg(\frac{2r^2}{r^2+a^2}+d-4\bigg)\bigg] \bigg\}\bar{R} = \tilde{T}\,,
    \label{radialeq_tortoise}
\end{align}
where $\tilde{T} \coloneqq \Delta_r r^{(d-4)/2} / (r^2+a^2)^{3/2} T$ and $\Delta^{\prime}_r=\dd{\Delta_r}/\dd{r}$, and $r^*$ is the tortoise coordinate defined by
\begin{align}\label{tortoise}
    \frac{\dd{r^*}}{\dd{r}}=\frac{r^2+a^2}{\Delta_r}\,.
\end{align}

When solving the radial equation \eqref{radialeq_tortoise}, the following boundary condition is imposed at the BH horizon ($r^* \to - \infty$) and at the cosmological horizon ($r^* \to \infty$):
\begin{align}\label{QNMBC}
    \bar{R}(r^*)\sim \begin{cases}
        e^{-i(\omega-m\Omega_{\text{h}})r^*}, & r^* \to -\infty \\
        e^{i(\omega-m\Omega_{\text{c}})r^*}, & r^* \to \infty.
    \end{cases}
\end{align}
This boundary condition is satisfied only when the frequency $\omega$ takes the discrete and complex values $\omega = \omega_{n} \in {\mathbb C}$, which is nothing but the QNM frequency. 
The index of $n = 0$, $1$, $2$, ... stands for the overtone number of QNMs.
The overtone number ``$n$'' is assigned in order of increasing damping rate.

The situation where a BH is embedded in de Sitter spacetime and the outer horizon coincides with the cosmological horizon is known as the Nariai limit. 
In the near-Nariai limit, $r_{\text{h}}\simeq r_{\text{c}}$ \cite{Ponglertsakul:2020ufm}, $\Delta_r$ and the surface gravity $\kappa_{\text{h}}$ are approximated as
\begin{align}
    \Delta_r &\simeq (d-1)(r_{\text{c}}-r)(r-r_{\text{h}})r^2_{\text{h}}\Lambda\,, \\ \kappa_{\text{h}} &\simeq \frac{(d-1)(r_{\text{c}}-r_{\text{h}})r^2_{\text{h}}\Lambda}{2(r^2_{\text{h}}+a^2)}\,.
\end{align}
We can express $r$ in terms of the tortoise coordinate in a closed form by integrating Eq.~(\ref{tortoise}):
\begin{align}
    r \simeq \frac{r_{\text{h}}+r_{\text{c}} e^{2\kappa_{\text{h}} r^*}}{1+e^{2\kappa_{\text{h}} r^*}}\,.
\end{align}
Substituting this relation into $\Delta_r$ gives
\begin{align}
    \Delta_r \simeq \frac{(d-1)(r_{\text{c}}-r_{\text{h}})^2r^2_{\text{h}}\Lambda}{4\cosh^2\left({\kappa_{\text{h}} r^*}\right)}.
\end{align}
Here we define the difference between the outer horizon $r_{\text{h}}$ and the cosmological horizon $h_c$ as $h \coloneqq r_{\text{c}}-r_{\text{h}}$, which serves as the expansion parameter from the Nariai limit. 
We consider a near-extremal case where $0<h \ll 1$.
Then, at the lowest order of $\mathcal{O}(h^2)$, the radial equation in Eq.~\eqref{radialeq_tortoise} becomes
\begin{align}
    \frac{d^2 \bar{R}}{d{r^*}^2} +& \qty[(\omega-m\Omega_{\text{h}})^2 - \frac{V_0}{\cosh^2{(\kappa_{\text{h}} r^*})}]\bar{R} = \tilde{T}, \label{Radialeq}\\
    V_0 \coloneqq& \frac{(d-1)h^2r^2_{\text{h}}\Lambda}{4(r^2_{\text{h}}+a^2)^2} \qty[\frac{j(j+d-5)a^2}{r^2_{\text{h}}} + \mu^2 r^2_{\text{h}} + A_{kjm}]\,,
    \label{V_0_formula}
\end{align}
where the term $V_0/\cosh^2{(\kappa_{\text{h}} r^*})$ is the PT potential \cite{Poschl1933, Ponglertsakul:2020ufm}.
This differential equation is known to be analytically solvable, and the solution is given as a superposition of hypergeometric functions \cite{Berti:2009kk, Ferrari:1984zz, Ponglertsakul:2020ufm}.

%%%%%%%%%%%%%%%%%%%%%%%%
\subsection{Angular Equation}
%%%%%%%%%%%%%%%%%%%%%%%%

Here, we review that the angular equation \eqref{Seq1} reduces the Heun's differential equation. 
In the limit of small rotation $(a\omega)\to 0$, the eigenvalue $A_{kjm}$ can be determined in the power series of $a \omega$, whose coefficients can be determined analytically \cite{Cho:2009wf, Berti:2005gp, Ponglertsakul:2020ufm}. 
The Heun's differential equation admits a three-term recurrence relation, which can be used to determine the expansion coefficients of the perturbation solution \cite{Ponglertsakul:2020ufm}.

By performing the variable transformation as $z=\cos^2 \theta$ and making the ansatz as
\begin{align}
    S(z) = 2^{\frac{m}{2}}(z-1)^{\frac{m}{2}}(2z)^{\frac{j}{2}}\bigg( z+\frac{1}{a^2\Lambda}\bigg)^{\frac{ia\omega}{2\sqrt{a^2\Lambda}}}H(z),
\end{align}
the angular equation \eqref{Seq1} can be written as the Heun's equation \cite{Ponglertsakul:2020ufm}:
\begin{align}
\begin{split}
    \frac{\dd^2{H}}{\dd{z}^2}&+\Bigg(\frac{\gamma}{z}+\frac{\delta}{z-1}+\frac{\epsilon}{z+\frac{1}{a^2\Lambda}}\Bigg)\frac{\dd{H}}{\dd{z}} \\
    &+\frac{\alpha\beta z-q}{z(z-1)\left(z+\frac{1}{a^2\Lambda}\right)}H=0,
\end{split}
\end{align}
where
\begin{align}
    \bar{g} &\coloneqq \frac{1}{4\Lambda}\left((d-1)\Lambda-\sqrt{\Lambda\left((d-1)^2\Lambda-4\mu^2\right) }\right),\\
    \alpha &\coloneqq \frac{1}{2}\left( j+m+\frac{ia\omega}{\sqrt{a^2\Lambda}}\right)+\bar{g}, \\
    \beta & \coloneqq \frac{1}{2}\left(j+d-1+m+\frac{ia\omega}{\sqrt{a^2\Lambda}} \right) -\bar{g}, \\
    \gamma &\coloneqq \frac{1}{2}(2j+d-3), \\
    \delta &\coloneqq 1+m, \\
    \epsilon &\coloneqq 1+\frac{ia\omega}{\sqrt{a^2\Lambda}}, \\
    q &\coloneqq \frac{m\omega}{2a\Lambda}+\frac{1}{4}\left(j+\frac{ia\omega}{\sqrt{a^2\Lambda}}\right)\left(j+d-3+\frac{ia\omega}{\sqrt{a^2\Lambda}}\right) \notag \\
    &+\frac{A_{kjm}-(j+m)(j+m+d-3)}{4a^2\Lambda}.
\end{align}
By expanding this Heun function using the hypergeometric functions, we can obtain a three-term recurrence relation for the expansion coefficients $\tilde{a}_p$ [see also Eq. (4.2) of \cite{Suzuki:1998vy}]:
\begin{align}
\begin{split}
    &\alpha+\beta-\epsilon-\gamma-\delta+1=0, \\
    &\alpha_0\tilde{a}_1+\beta_0\tilde{a}_0 = 0,\\
    &\alpha_p\tilde{a}_{p+1}+\beta_p\tilde{a}_p+\gamma_p\tilde{a}_{p-1}=0,\; (p=1,2,\ldots),
\end{split}
\end{align}
where
\begin{align}
    \alpha_p & \coloneqq-\frac{(p+1)(p+\tilde{r}-\alpha+1)(p+\tilde{r}-\beta+1)(p+\delta)}{(2 p+\tilde{r}+2)(2 p+\tilde{r}+1)}, \\
    \beta_p & \coloneqq\frac{1}{(2 p+\tilde{r}+1)(2 p+\tilde{r}-1)}[p(p+\tilde{r})(\gamma-\delta) \epsilon \notag\\ 
    &+(p(p+\tilde{r})+\alpha \beta)(2 p(p+\tilde{r})+\gamma(\tilde{r}-1))]\\
    &-\frac{p(p+\tilde{r})}{a^2 \Lambda}-q, \notag\\
    \gamma_p &\coloneqq -\frac{(p+\alpha-1)(p+\beta-1)(p+\gamma-1)(p+\tilde{r}-1)}{(2 p+\tilde{r}-2)(2 p+\tilde{r}-1)},\\
    \tilde{r}&\coloneqq \alpha+\beta-\epsilon=\gamma+\delta-1= j+m+\frac{d-3}{2} .
\end{align}
We can finally obtain a single continued fraction equation:
\begin{align}\label{CFE}
    \beta_0-\frac{\alpha_0 \gamma_1}{\beta_1-} \frac{\alpha_1 \gamma_2}{\beta_2-} \frac{\alpha_2 \gamma_3}{\beta_3-} \ldots=0.
\end{align}
Solving this equation, the eigenvalue $A_{kjm}$ is obtained \cite{Leaver:1985ax}.
When the spin parameter is zero, $a = 0$, the series expansion becomes divergent unless requiring Eq. (\ref{CFE}) to be truncated at some finite terms \cite{Leaver:1985ax}. 
Consequently, imposing $\beta_k = 0$ for some integer $k\ge0$ gives us the eigenvalue of the separation constant in the analytical form $A_{kjm}=(2k+j+m)(2k+j+m+d-3)$ \cite{Berti:2005gp, Kodama:2009rq}, which is reminiscent of the derivation of the Hermite polynomials in the quantization of a harmonic oscillator.

When $a=0$, we can directly specify this quantum number $k$ to get the separation constant. 
However, when $a\ne0$, we can obtain $A_{kjm}$ numerically by using Newton's method to find the root of the continued fraction equation.

%%%%%%%%%%%%%%%%%%%%%%%%
\section{Mode excitation in the Nariai limit}
\label{Sec_mode_excitation}
%%%%%%%%%%%%%%%%%%%%%%%%

We use the Green's function technique to solve the radial Eq. (\ref{Radialeq}). 
To construct the Green's function, two independent homogeneous solutions, $\bar{R}_{\rm in}$ and $\bar{R}_{\rm out}$, are necessary.
The in-mode solution $\bar{R}_{\rm in}$ satisfies the following boundary condition:
\begin{align}
    \bar{R}_{\rm in} \rightarrow 
    \begin{cases} 
    e^{-i(\omega-m\Omega_{\text{h}}) r^*} & \left(r^* \rightarrow-\infty\right)\,, \\ 
    A_{\text{out}}(\omega)e^{i(\omega-m\Omega_{\rm c}) r^*} \\ \; + A_{\text{in}}(\omega)e^{-i(\omega-m\Omega_{\rm c}) r^*} & \left(r^* \rightarrow+\infty\right)\,,
    \end{cases}
    \label{INmode_boundary}
\end{align}
where the event horizon and the cosmological horizon are approximated to share the same horizon velocity $\Omega_{\rm h}$ in the Nariai limit. 
The homogeneous solutions can be expressed in terms of the hypergeometric function \cite{Berti:2009kk, Kuntz:2025gdq}. 
By virtue of this, $A_{\text{in}}$ and $A_{\text{out}}$ have an analytic form involving the gamma function \cite{Ferrari:1984zz, Berti:2006wq, Berti:2009kk}:
\begin{align}
\begin{split}
    A_{\text{out}} &= \frac{\Gamma(1-i\omega/\kappa_{\text{h}})\Gamma(i\omega/\kappa_{\text{h}})}{\Gamma(1+\beta)\Gamma(-\beta)}\,, \\
    A_{\text{in}} &= \frac{\Gamma(1-i\omega/\kappa_{\text{h}})\Gamma(-i\omega/\kappa_{\text{h}})}{\Gamma(1+\beta-i\omega/\kappa_{\text{h}})\Gamma(-\beta-i\omega/\kappa_{\text{h}})}\,,
\end{split}
\label{ain_aout}
\end{align}
where 
\begin{equation}
\beta=-1/2+i\sqrt{V_0/\kappa_{\text{h}}^2-1/4}\,.
\label{beta_definition}
\end{equation}
Imposing the QNM boundary condition, i.e., $A_{\text{in}} (\omega_{n}) = 0$ [see Eq.~\eqref{QNMBC} and Eq.~\eqref{INmode_boundary}], we obtain the analytic formula for QNM frequencies $\omega_{n}$:
\begin{align}\label{anaQNM}
    \omega_{n}^{\pm} &= m\Omega_{\text{h}} + \kappa_{\text{h}} \qty[\pm\sqrt{\frac{V_0^\pm}{\kappa_{\text{h}}^2}-\frac{1}{4}}-i\qty(n+\frac{1}{2})]\,,
\end{align}
where the superscript, positive and negative sign, stands for the prograde and retrograde modes, respectively. Note that the separation constant $A_{kjm}^\pm$ in $V_0$ [Eq.~\eqref{V_0_formula}] depends on whether the mode is prograde or retrograde.

When the initial source is located in the far region in the tortoise coordinate $(r^* \gg 1)$, the time-domain ringdown waveform measured at the far region can be approximated by\footnote{The factor $\bar{R}_{\rm out} \simeq e^{i (\omega-m \Omega_{\rm c}) r^*}$ is factored out from the integrand for simplicity as it does not affect our discussion.}
\begin{equation}
    \bar{R} (t) \sim \frac{1}{2 \pi} \int \dd \omega \frac{\tilde{T} (\omega)}{2i (\omega - m \Omega_{\rm c})} \tilde{X}_{\rm G} (\omega) e^{-i \omega t}\,,
\end{equation}
where $\tilde{T} (\omega)$ is the source term, and $\tilde{X}_{\rm G}$ is defined as
\begin{align}
    \tilde{X}_{\text{G}}(\omega) \coloneqq \frac{A_{\text{out}} (\omega)}{A_{\text{in}} (\omega)}.
\end{align}
The poles of the Green's function, i.e., QNMs, are nothing but the zeros of $A_{\rm in} (\omega)$.
It means that the excitation of QNMs is encoded in the spectral function $\tilde{X}_{\rm G}$.
Also, for general source term $\tilde{T}$, one can decompose the original waveform $\bar{R}$ into the two time-domain waveoforms:
\begin{equation}
\bar{R} = h_{\rm G} (t) * h_{\rm T} (t)\,,
\end{equation}
where $*$ stands for the convolution:
\begin{equation}
\bar{x} (t) * \bar{y} (t) \coloneqq \int_{- \infty}^{\infty} \dd \tau \bar{x} (\tau) \bar{y} (t-\tau)\,,
\end{equation}
and
\begin{align}\label{hG}
    h_{\rm G} &\coloneqq \frac{1}{2\pi} \int\dd{\omega}\tilde{X}_G(\omega) e^{-i\omega t}\,,\\
    h_{\rm T}  &\coloneqq \frac{1}{2\pi} \int\dd{\omega}\tilde{T}(\omega) e^{-i\omega t}\,.
\end{align}
In the following, we consider the scalar QNM excitation of $h_{\rm G}$ to examine the stability of the BH in the Nariai limit.
Once we understand the QNM excitation in $h_{\rm G}$, the QNM amplitude in the original ringdown waveform $\bar{R}$ can be read by convoluting with a desired source $h_{\rm T}$.

Since $A_{\text{in}}$ and $A_{\text{out}}$ can be written analytically as Eq.~\eqref{ain_aout}, the excitation factor\footnote{The conventional definition of the excitation factor is $A_{\rm out} (\omega_n) / [2 i \omega (\dd A_{\rm in} / \dd \omega)]_{\omega = \omega_n}$.}
\begin{align}\label{EFs}
    B_n^{\pm} = A_{\text{out}}\qty(\frac{dA_{\text{in}}}{d\omega})^{-1} \bigg|_{\omega=\omega_n^{\pm}},
\end{align}
can also be determined analytically:
\begin{align}
\begin{split}
    B_n^{+} &= \frac{i(-1)^n\kappa_{\text{h}}}{n!}\frac{\Gamma(-n-2\beta-1)\Gamma(n+\beta+1)}{\Gamma(-n-\beta-1)\Gamma(-\beta)\Gamma(\beta+1)},\\
    B_n^{-} &= \frac{i(-1)^n\kappa_{\text{h}}}{n!}\frac{\Gamma(n-\beta)\Gamma(1+2\beta-n)}{\Gamma(\beta-n)\Gamma(-\beta)\Gamma(\beta+1)}.
\end{split}
\label{PT_excitation_fac_beta}
\end{align}
As the excitation factor is associated with the residue of the pole in $\tilde{X}_{\rm G}$, i.e., the amplitude of each QNM in $h_{\rm G}$, the waveform $h_{\rm G}$ can be reconstructed with the superposition of QNMs with amplitude $B_n^{\pm}$ when $A_{\text{in}}$ has only simple poles \cite{Oshita:2024wgt}:
\begin{align}\label{hE}
    h_{\text{E}} = -i\sum_{n=0}^{N_+}B_n^{+} e^{-i\omega_n^+t}-i\sum_{n=0}^{N_-}B_n^{-} e^{-i\omega_n^-t},
\end{align}
where $N_{\pm}$ should be ideally set to $N_{\pm} = \infty$ to exactly reconstruct the waveform $h_{\rm G}$.
Practically, one can truncate at finite but large overtone numbers, $N_{\pm} < \infty$, to approximately reconstruct the waveform \cite{Oshita:2024wgt}.
Note that it was shown \cite{Casals:2009zh} that the PT potential does not have a branch point, leading to the power-law tail.

When the square root in Eq.~(\ref{anaQNM}) vanishes, the double-pole QNMs appear as $\omega_n^+ = \omega_n^-$ whose BH parameters are an EP.
As the BH parameters approach the EP, the excitation factors of QNMs that become degenerate at the EP are significantly enhanced \cite{Motohashi:2024fwt}. 
However, it was pointed out \cite{Oshita:2025ibu} that the apparent amplification does not occur due to their destructive excitations.
On the other hand, the transient linear growth, which is a unique feature in the QNM excitation near or at the EP, may happen at the early stage of QNM excitation \cite{Yang:2025dbn}.
Therefore, the response of spacetime near the EP becomes a crucial probe for examining the EP or AC appearing as a consequence of the non-Hermitian nature of BHs.

In the following, we analytically investigate the characteristics of QNM excitation near EPs: i) transient linear growth in QNM amplitude, ii) stability or destructive excitation of QNMs near EPs, in the Nariai limit with $d=4$ (main text) and $d=5$ (Appendix~\ref{App_d5_case}).
We also show that an EL exists in the parameter space spanned by the BH spin $a$ and mass of the scalar field $\mu$.

We here consider the QNM excitation around EPs causing the degeneracy between the prograde and retrograde modes with the same overtone number $n$, i.e., $\omega_n^+ = \omega_n^-$. 
For any non-zero values of $k$, $j$, or $m$, no EP is found within the parameter space we investigated in this work. 
As such, we study the massive scalar mode with $k=j=m=0$. 
As the perturbation is axisymmetric for $m=0$, QNM frequencies exhibit the symmetry of $\omega_n^{-}=-(\omega_n^{+})^*$, and the eigenvalue of the separation constant satisfies $A_{kjm}^{-}=(A_{kjm}^{+})^*$.

%%%%%%%%%%%%%%%%%%%%%%%%
\subsection{Transient linear growth of QNM excitation near or at an EP}
%%%%%%%%%%%%%%%%%%%%%%%%
In Ref.~\cite{Yang:2025dbn}, it was shown that the excitation of nearly double-pole QNMs exhibits linear growth at the early stage of ringdown, and is exponentially suppressed due to their QNM damping afterwards.
They interpreted the behavior as the beating phenomenon arising from the interference of two QNMs with closely spaced complex frequencies.
However, this behavior can be reproduced even when we consider the double-pole QNM, i.e., exact degeneracy of QNMs.
Let us consider the excitation of two QNMs with $\omega =  \omega_{\rm EP} (\eqqcolon \Omega_a)$ and $\omega = \omega_{\rm EP} + \delta \omega (\eqqcolon \Omega_b)$.
When the two QNMs are close to each other, the function $A_{\rm in} (\omega)$ around the two QNM frequencies can be written as
\begin{equation}
A_{\rm in} = f(\omega) (\omega - \Omega_a) (\omega - \Omega_b)\,,
\end{equation}
where $f(\omega)$ is regular around the two QNM frequencies.
In this case, the two QNM excitations in the ringdown waveform $h_{\rm G}$ can be written as
\begin{align}
\begin{split}
    h_{\rm near-EP} &\coloneqq -i \frac{A_{\rm out} (\Omega_b)}{f(\Omega_b) (\Omega_b - \Omega_a)} e^{-i \Omega_b t} \\
    &-i \frac{A_{\rm out} (\Omega_a)}{f(\Omega_a) (\Omega_a - \Omega_b)} e^{-i \Omega_a t}\,.
\end{split}
\label{h_nearEP_original}
\end{align}
Expanding this with respect to $\delta \omega = \Omega_b - \Omega_a$, one obtains
\begin{align}
\begin{split}
    h_{\rm near-EP} &= -i\frac{e^{-i \omega_{\rm EP} t}}{f(\omega_{\rm EP})^2}\left[(A^\prime_\text{out}-itA_\text{out})f - A_\text{out}f^\prime\right]_{\omega = \omega_{\rm EP}}\\
    &+ \mathcal{O}(\delta \omega)\,,\\
    &= -\left( \frac{A_\text{out}}{f}\right)_{\omega=\omega_{\rm EP}}te^{-i \omega_{\rm EP} t} \\
    & -i \frac{d}{d\omega}\left( \frac{A_\text{out}}{f}\right)_{\omega=\omega_{\rm EP}}e^{-i \omega_{\rm EP} t} + \mathcal{O}(\delta \omega)\,,
    \end{split}
    \label{h_nearEP}
\end{align}
whose second term is consistent with Eq.~(10) in \cite{Oshita:2025ibu}.
On the other hand, for the double-pole QNM, i.e., $\delta \omega = 0$, the QNM excitation is obtained from the residue of a double-pole at $\omega = \omega_{\rm EP}$ in $A_{\rm out} / A_{\rm in} e^{-i \omega t}$:
\begin{align}
\begin{split}
    h_{\rm EP} &= -i \frac{d}{d \omega} \left[ (\omega - \omega_{\rm EP})^2 \frac{A_{\rm out}(\omega)}{f(\omega) (\omega - \omega_{\rm EP})^2} e^{-i \omega t} \right]_{\omega = \omega_{\rm EP}}\,,\\
    &= -i\frac{e^{-i\omega_{\rm EP} t}}{f(\omega_{\rm EP})^2}\left[(A^\prime_\text{out}-itA_\text{out})f - A_\text{out}f^\prime\right]_{\omega = \omega_{\rm EP}}.
    \label{h_EP}
    \end{split}
\end{align}
Indeed, this is expected, and one can find that the QNM amplitudes, $h_{\rm near-EP}$ in \eqref{h_nearEP} and $h_{\rm EP}$ in \eqref{h_EP}, are the same, $\displaystyle \lim_{\delta \omega \to 0} h_{\rm near-EP} = h_{\rm EP}$.

As we presented previously, the Green's function in the Nariai limit can be written in an analytic way, and QNM frequencies and excitation factors have closed forms.
This is advantageous to explore the structure of EP in the BH parameter space over the mass of the scalar field $\mu$ and the spin parameter $a$.

The mode degeneracy between $\omega_{n}^+$ and $\omega_{n}^-$ occurs at $\beta=-1/2$ [c.f. Eqs.~\eqref{beta_definition} and \eqref{anaQNM} (see also Eq.~\eqref{sqrtzero} for the spinning case)]. 
Then, we can determine the residues of double poles at $\omega_n^+=\omega_n^-= -i\kappa_{\text{h}}(n+1/2)\eqqcolon\omega_{\rm EP}$ in $h_{\rm G}$ \eqref{hG} as follows \cite{Abramowitz:1964}:
\begin{align}
\begin{split}
    &\mathrm{Res}_{\omega=\omega_{\rm EP}} \qty[\frac{A_{\text{out}}}{A_{\text{in}}}e^{-i\omega t}] \\
    &= i\kappa_{\text{h}}g_n^{(1)} \frac{\Gamma(n+1/2)}{\pi\Gamma(-n-1/2)}e^{-\kappa_{\text{h}}(n+1/2)t} \\
    &\times\qty[\kappa_{\text{h}}g_n^{(1)}t-2g_n^{(1)}\qty(\psi^{(0)}(n+1/2)+\frac{1}{2n+1})+2g_n^{(0)}],
    \end{split}
\end{align}
where we used \eqref{ain_aout} and Polygamma function $\psi^{(0)}(z):=\Gamma^\prime(z)/\Gamma(z)$:
\begin{align}
    \psi^{(0)}(n+1/2) &= \begin{cases}
        -\gamma_\text{Euler}-2\ln2 & n=0, \\
        -\gamma_\text{Euler}-2\ln2 + 2 \sum_{k=1}^{n}\frac{1}{2k-1} & n\ge1,
    \end{cases} \\
    \gamma_\text{Euler} &\coloneqq \lim_{n\to\infty}\qty(\sum_{k=1}^{n}\frac{1}{k}-\ln n).
\end{align}
The constants $g_n^{(1)}$ and $g_n^{(0)}$ are defined by the expansion coefficients of the Gamma function $\Gamma (x)$ around its poles $x = -n\in\mathbb{Z}_{\le0}$:
\begin{align}
\begin{split}
    \Gamma(-n+\varepsilon) &= \frac{(-1)^n}{n!\varepsilon}+\frac{(-1)^n(H_n-\gamma_\text{Euler})}{n!} + \mathcal{O}(\varepsilon) \\
    &\equiv\frac{g_n^{(1)}}{\varepsilon} + g_n^{(0)} + \mathcal{O}(\varepsilon),
\end{split}
\end{align}
where
\begin{align}
    H_n \coloneqq \begin{cases}
        0 & n=0, \\
        \sum_{k=1}^n \frac{1}{k} & n \ge1.
    \end{cases}
\end{align}
In the case of $a=0$ (non-spinning BH), the eigenvalue of the separation constant $A_{kjm}$ is independent of the overtone number.
In such a case, the EP condition $V_0/\kappa_h - 1/4$ is also independent of the overtone number [see Eq.~\eqref{V_0_formula}].
Therefore, the EP occurs for all $n$ at the same parameter.
We then obtain the superposed QNM amplitude, $h_\text{E}$, as
\begin{align}\label{hEa=0}
\begin{split}
    h_{\text{E}} &= \frac{\kappa_{\text{h}}}{\pi} \sum_{n=0}^{\infty} g_n^{(1)} \frac{\Gamma(n+1/2)}{\Gamma(-n-1/2)}e^{-\kappa_{\text{h}}(n+1/2)t} \\
    &\times\qty[\kappa_{\text{h}}g_n^{(1)}t-2g_n^{(1)}\qty(\psi^{(0)}(n+1/2)+\frac{1}{2n+1})+2g_n^{(0)}]\,.
\end{split}
\end{align}

\begin{figure*}[t]
  \centering
  \includegraphics[width=\linewidth, trim=0cm 0cm 0cm 0cm]{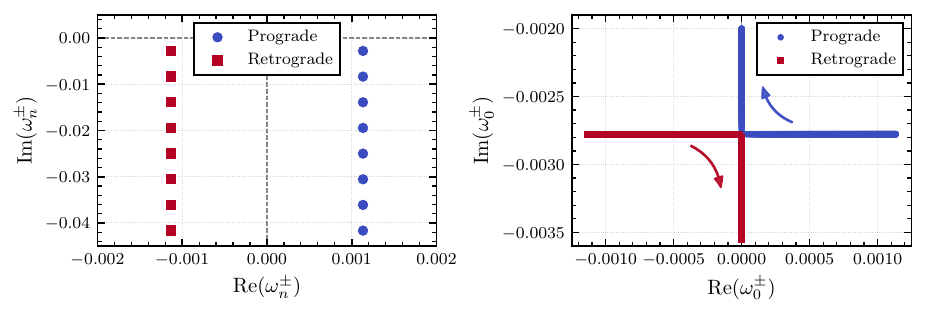}
  \caption{
    ({\bf Left}) Distribution of QNMs for prograde and retrograde modes when $a=0$, $\mu=0.18$, and $k=j=m=0$ up to $n=7$ modes.
    ({\bf Right}) Trajectories of the fundamental QNMs, $\omega_0^{\pm}$, near or at the exceptional point. 
    The value $\mu$ decreases from $0.18$ to $0.16$ along the arrows, and the two QNMs intersect at $\mu=1/6$.
  }
  \label{fig:qnm_combined4}
\end{figure*}

On the other hand, for a spinning BH with $a>0$, the eigenvalue $A_{kjm}$ depends on the overtone number $n$ in general, and an EP appears only for a single overtone number, say $n = n_{\rm EP}$.
We then have the following waveform:
\begin{align}
\begin{split}
    h_{\text{E}} &= -i\sum_{n=0, n\ne n_{\rm EP}}^{\infty} \left(B_n^+ e^{-i\omega_n^+t} + B_n^{-} e^{-i\omega_n^-t}\right) \\
    &+\frac{\kappa_{\text{h}}}{\pi}  g_{n_{\rm EP}}^{(1)} \frac{\Gamma(n_{\rm EP}+1/2)}{\Gamma(-n_{\rm EP}-1/2)}e^{-\kappa_{\text{h}}(n_{\rm EP}+1/2)t} \bigg[\kappa_{\text{h}}g_{n_{\rm EP}}^{(1)}t \\
    &-2g_{n_{\rm EP}}^{(1)}\qty(\psi^{(0)}(n_{\rm EP}+1/2)+\frac{1}{2n_{\rm EP}+1})+2g_{n_{\rm EP}}^{(0)} \bigg].
\end{split}
\end{align}
This expression contains a term proportional to $te^{-i\omega_n^{\pm}t}$, which is the interference pattern unique to the (near) double-pole QNM \cite{Yang:2025dbn}. 
Our results demonstrated that the massive scalar field in the Nariai limit also exhibits the excitation of double-pole QNM and that the scattering wave in the PT barrier serves as a simple and analytic toy model for the QNM excitation at an EP.

%%%%%%%%%%%%%%%%%%%%%%%%
\subsection{Stability of time-domain ringdown and instability of QNM amplitude}
%%%%%%%%%%%%%%%%%%%%%%%%

We discuss here the stability of ringdown waveforms near the EP parameter.
This is a demonstration of the result in Ref.~\cite{Oshita:2025ibu} by using the analytic perturbation solution in the Nariai limit.
Our result presented in this subsection is well consistent with that in Ref.~\cite{Oshita:2025ibu}.
That is, we confirm that the ringdown waveform constructed with the superposed QNMs is insensitive to the BH parameters near EP, whereas the individual amplitude of each QNM is quite sensitive to the parameters.
Therefore, the amplification of QNM near the EP parameter is an apparent amplification, and the observable (i.e., superposed QNMs or ringdown waveform) is still stable and insensitive to EP.
This is very similar to the discussion of the QNM instability and ringdown stability \cite{Nollert:1996rf,Barausse:2014tra,Jaramillo:2020tuu,Jaramillo:2021tmt,Cheung:2021bol,Berti:2022xfj}.

Let us consider the non-spinning BH $a=0$ in the four dimensions $d =4$.
To take the Nariai limit, one has to find the condition for the cosmological constant $\Lambda = \Lambda(a)$ which satisfies $r_{\text{h}}=r_{\text{c}}$. 
For $a=0$, it is $\Lambda=1/27$ and $r_{\text{h}}=r_{\text{c}}=3$.
The quantities relevant to the following analysis can be summarized as
\begin{align}
    \kappa_{\text{h}} &\simeq \frac{h}{18}, \\
    V_0 &\simeq \frac{h^2}{324}(4\mu^2+A_{kjm}), \\
    \omega_n &\simeq \pm\frac{h}{18}\sqrt{9\mu^2+A_{kjm}-\frac{1}{4}}-\frac{hi}{18}(n+\frac{1}{2}), \label{PTQNMd4} \\
    A_{kjm}&=(2k+m)(2k+m+1). \label{SCd4}
\end{align}
The high-dimensional angular space, $\varphi_j$, is absent for $d=4$, and we have $Y=1$ and $j=0$ in Eq.~(\ref{LaplaceY}). 
The parameter $h$ serves as the expansion parameter of the Nariai limit. We here take the value of $h=0.1$.
In this case, the QNM distribution has the homogeneous separation along the imaginary axis, which can be captured analytically (left panel in FIG.~\ref{fig:qnm_combined4}).

We find that for $k=j=m=0$ (or monopole perturbation), we can make the term under the square root in Eq.~(\ref{PTQNMd4}) zero by introducing and tuning the other degree of freedom, i.e., mass of the scalar field $\mu$ (right panel in FIG. \ref{fig:qnm_combined4}).
We find that the EP arises when $\mu = 1/6$.
For $a=0$, the eigenvalue $A_{kjm}$ in Eq.~(\ref{SCd4}) is independent of the overtone number $n$, and the prograde and retrograde modes for all overtones degenerate at the same mass $\mu=1/6$.

\begin{figure*}[!thpb]
  \centering

  % --- 1段目 ---
  \includegraphics[width=0.49\textwidth]{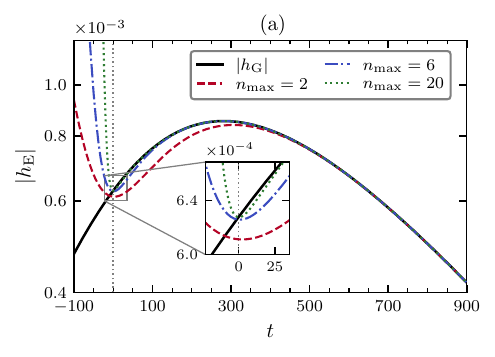}
  \hfill
  \includegraphics[width=0.49\textwidth]{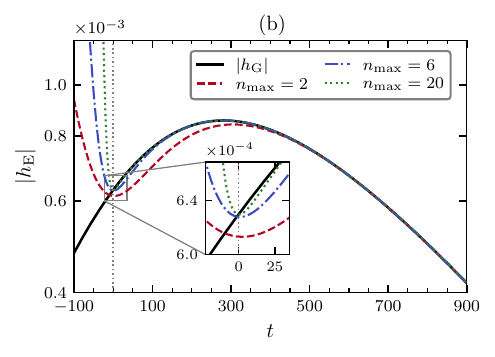}

  \vspace{1em}

  % --- 2段目 ---
  \includegraphics[width=0.49\textwidth]{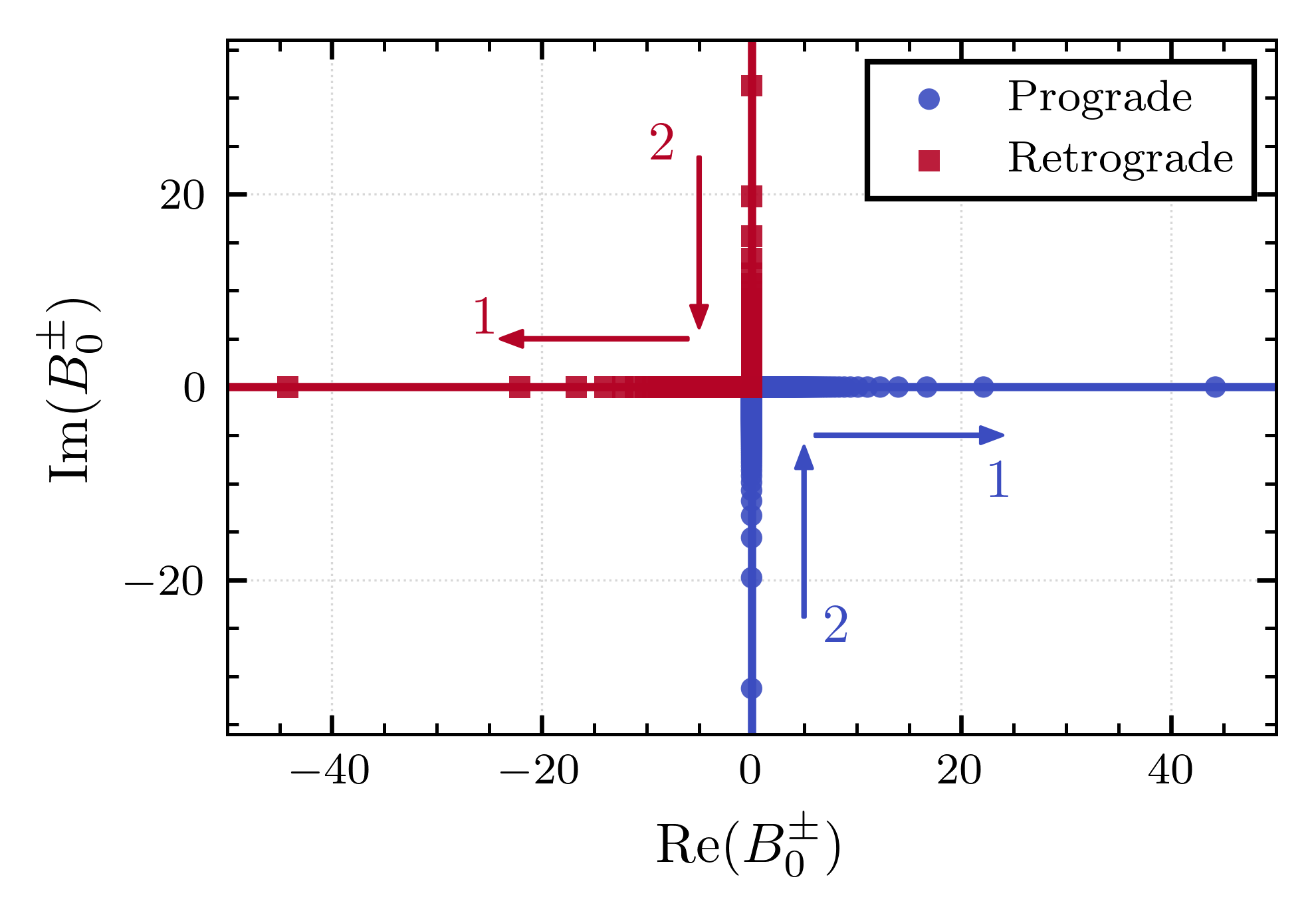}%
  \hfill
  \includegraphics[width=0.49\textwidth]{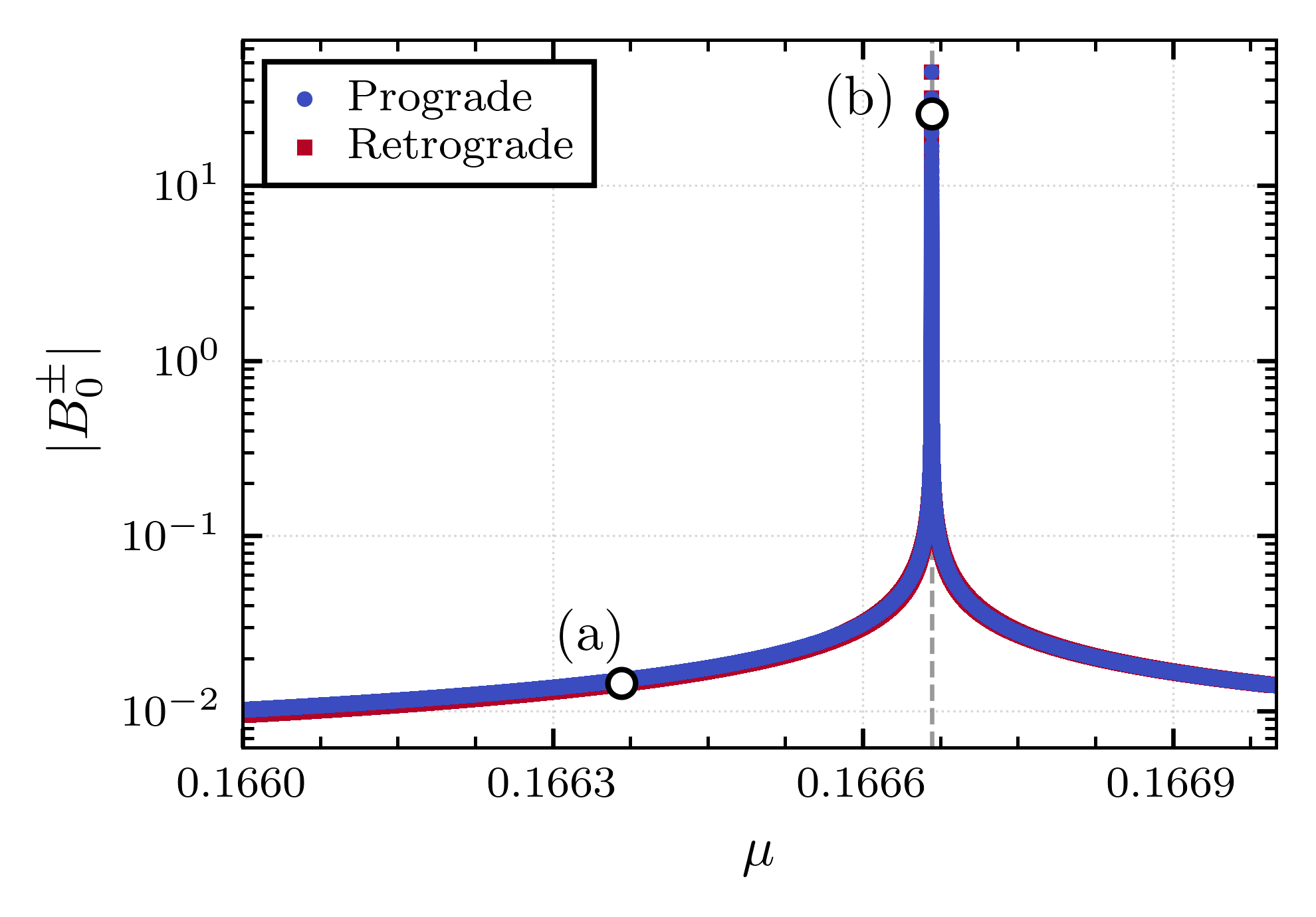}

  \caption{
    ({\bf Upper panels}) Ringdown waveform $|h_{\rm G}|$ (black solid) and the ringdown reconstruction $|h_{\rm E}|$ with $n_{\rm max} = 2$ (red dashed), $n_{\rm max} = 6$ (blue dot-dashed) and $n_{\rm max} = 20$ (green dotted) for (a) $\mu=1/6-3\times10^{-4}$ and (b) $\mu=1/6-10^{-10}$, where $N_+ = N_- \eqcolon n_{\rm max}$.
    ({\bf Bottom left}) The excitation factors $B_0^{\pm}$ (squares and circles) are calculated with the range of $0.166\leq \mu \leq 0.167$ in the step size of $10^{-10}$.
    ({\bf Bottom right}) The absolute values of the excitation factor $|B_0^{\pm}|$ of both the prograde and retrograde mode.
    The excitation factor diverges at $\mu=1/6$ (gray-dashed line), where the double-pole QNM appears.
  }
  \label{fig:4plots1}
\end{figure*}

When two QNMs degenerate in the complex frequency plane, it has been shown that the excitation factors of the modes diverge (for an analytic discussion on the divergence, see Appendix~\ref{Sec_div_EFs}).
We calculate the excitation factors around the EP and confirm that they exhibit a divergent behavior at the same mass $\mu=1/6$ where EPs are achieved (two bottom panels in FIG.~\ref{fig:4plots1}).

To examine the stability of the waveform near the EPs where the excitation factors diverge, we select the two parameters (a) $\mu=1/6-3\times10^{-4}$ and (b) $\mu=1/6-10^{-10}$, which are sufficiently close to $\mu=1/6$ but differ by an order of magnitude in the absolute value of the excitation factors.
For the two cases, (a) and (b), we perform the ringdown reconstruction by using Eq.~(\ref{hE}). 
The two upper panels in FIG.~\ref{fig:4plots1} show the waveforms in the case of (a) and (b). 
They demonstrate that the waveforms remain stable even though the excitation factors are different by the order of magnitude $\sim 10^3$, implying that the EP of QNMs does not contribute to the instability of the Nariai BH.

As the two modes approach each other, the excitation factors are amplified and diverge at EP, but their phase is opposite, $|\text{Arg}(E_{0}^+ / E_{0}^-)| \simeq \pi$ (see Ref.~\cite{Oshita:2025ibu}). 
As a result, we find that the superposed QNMs remain finite, which is consistent with the previous research \cite{Oshita:2025ibu}.

Note that each QNM has a significantly small Q value, defined as $|\text{Re}(\omega_n^{\pm}) / \text{Im}(\omega_n^{\pm})|$, which means each mode is an over-damping mode.
Nevertheless, the ringdown $h_{\rm G}$ in the upper panels in FIG.~\ref{fig:4plots1} exhibits a growing phase at the early ringdown.
This transient linear growth is a footprint of the excitation of nearly double-pole QNMs.
One can reconstruct the linear growth in $h_{\rm G}$ by summing up overtone contribution \eqref{hE} with $N_+ = N_- =20$ (green dotted line in the upper panels in FIG.~\ref{fig:4plots1}).

\begin{figure}[t]
\centering
\includegraphics[width=\columnwidth]{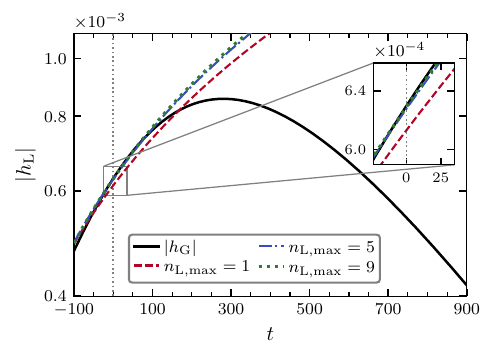}
\caption{Transient linear growth at the EP. 
$h_\text{G}$ is calculated with the parameter $\mu=1/6$ (black solid), which is the EP parameter. 
The red dashed line, blue dash-dotted line, and green dotted line are the amplitude factors, $h_{\text{L}}$ in Eq.~\eqref{hLa=0} with $n_{\text{L,max}}=1$, $5$, and $9$, respectively.}
\label{Beatingd4}
\end{figure}

This linear growth in $h_{\rm G}$, which is obtained by the numerical integration in \eqref{hG}, agrees with the following amplitude factor:
\begin{align}\label{hLa=0}
\begin{split}
    h_\text{L} &= \frac{\kappa_{\text{h}}}{\pi} \sum_{n=0}^{n_{\text{L,max}}} g_n^{(1)} \frac{\Gamma(n+1/2)}{\Gamma(-n-1/2)}\\
    &\times\qty[\kappa_{\text{h}}g_n^{(1)}t-2g_n^{(1)}\qty(\psi^{(0)}(n+1/2)+\frac{1}{2n+1})+2g_n^{(0)}]\,,
\end{split}
\end{align}
which is analytically obtained by factoring out the QNM function, $e^{-i \omega_n^{\pm} t}$, from Eq.~\eqref{hEa=0}. 
FIG.~\ref{Beatingd4} shows the waveforms observed at the EP parameters $\mu=1/6$, together with the plots of $h_\text{L}$ with $n_\text{L,max}=1, 5, 9$. 
From the comparison, we confirm that the growing amplitude is well consistent with the theoretical prediction of the transient linear growth caused by the double-pole QNMs. 
The emergence of the linear growth near an EP parameter was predicted in \cite{Yang:2025dbn}. 
The waveform model to verify the linear growth of QNM excitation near or at an EP was recently studied in \cite{PanossoMacedo:2025xnf}.

%%%%%%%%%%%%%%%%%%%%%%%%
\subsection{Exceptional line: BH spin and scalar mass}
\label{Subsec_d=4_a>0}
%%%%%%%%%%%%%%%%%%%%%%%%
Let us consider a spinning BH, $a \neq 0$, and study the QNM distribution in the $a$-$\mu$ parameter space.
In this subsection, we show that the Kerr-de Sitter BH with a massive scalar field in the Nariai limit exhibits an EL.

Introducing a non-zero spin parameter, a numerical procedure is involved to obtain the eigenvalues $\omega_{n}^{\pm}$ and $A_{n}^{\pm}$.
Those values can be obtained by simultaneously solving the radial equation Eq.~\eqref{Radialeq} and angular equation Eq.~\eqref{Seq1}.
As an example, we show the distribution of QNMs for the parameter set of $a=0.5$ and $\mu=0.18$ in Table.~\ref{tab:qnm_prograde}. 
We find that the value of $|\text{Re}(\omega_{n}^{\pm})|$ gradually decreases as the overtone number increases, which does not mean that one can no longer use the simple formula of QNMs in \eqref{anaQNM}. It is the consequence of the dependence of $V_0$ on $\omega$ via $A_n^{\pm}$. 

\begin{table}[htpb]
    \centering
    \caption{Prograde QNM frequencies for $a=0.5$, $\mu=0.18$, and $k=j=m=0$ up to $n=4$. 
The absolute values of the real part of the QNMs gradually decrease as the overtone number increases.
Retrograde QNM frequencies are obtained with $\omega_n^- = - (\omega_n^+)^*$.
}
    \label{tab:qnm_prograde}
    \begin{tabular}{c c c}
        \hline \midrule
        $n$ & $\mathrm{Re}(\omega_n^{+})$ & $\mathrm{Im}(\omega_n^{+})$ \\
        \midrule
        0 & $8.846 \times 10^{-4}$ & $-2.860 \times 10^{-3}$ \\
        1 & $8.844 \times 10^{-4}$ & $-8.579 \times 10^{-3}$ \\
        2 & $8.841 \times 10^{-4}$ & $-1.430 \times 10^{-2}$ \\
        3 & $8.836 \times 10^{-4}$ & $-2.002 \times 10^{-2}$ \\
        4 & $8.830 \times 10^{-4}$ & $-2.574 \times 10^{-2}$ \\
        \hline \hline
    \end{tabular}
\end{table}

\begin{figure*}[t]
    \centering
    \begin{minipage}{0.48\textwidth}
        \centering
        \includegraphics[width=\linewidth]{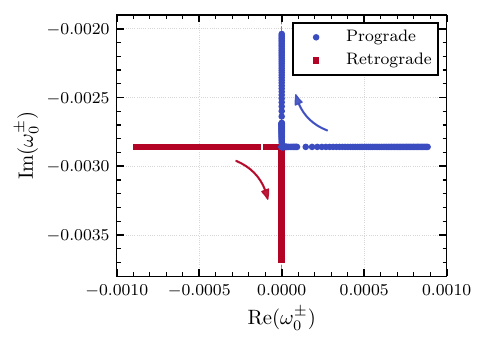}
    \end{minipage}
    \begin{minipage}{0.48\textwidth}
        \centering
        \includegraphics[width=\linewidth]{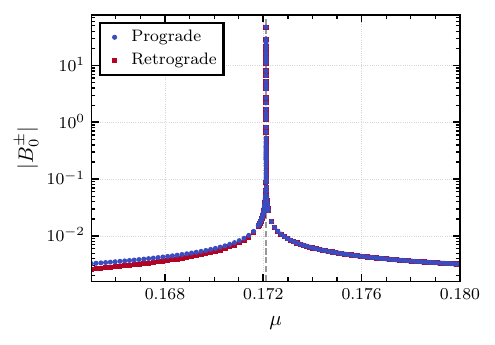}
    \end{minipage}

    \caption{
    ({\bf Left}) The distribution of the fundamental QNMs at $a=0.5$ and $0.165 \leq \mu \leq 0.18$. 
    The existence of EP is confirmed even for $a>0$. 
    The arrows indicate the direction along which $\mu$ decreases from $0.18$ to $0.165$. 
    For the fundamental mode, the EP is found at $\mu \simeq 0.17211$.
    ({\bf Right}) The absolute values of the excitation factor $B_0^{\pm}$ for $a=0.5$. 
    It can be seen that the excitation factor diverges at the EP parameter $\mu \simeq 0.17211$ (gray-dashed line).
    }
    \label{QNM_EF_a>0}
\end{figure*}

The left panel in FIG.~\ref{QNM_EF_a>0} shows the values of $\omega_{n}^{\pm}$ in the range of $0.165 \leq \mu \leq 0.18$ for $a=0.5$.
We find that for $k=j=m=0$, double-pole QNM arises on the imaginary axis for $a = 0.5$ and $\mu \simeq 0.17211$.
We also calculate the excitation factors for the prograde and retrograde modes for various values of $\mu$ (FIG.~\ref{QNM_EF_a>0}). 

\begin{figure}[!t]
\centering
\includegraphics[width=0.9\columnwidth]{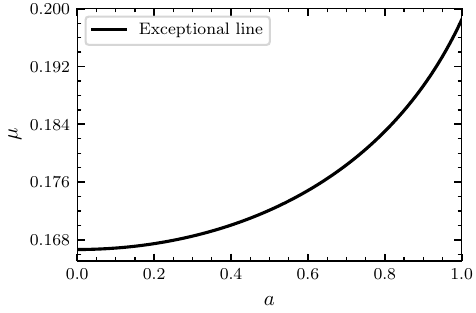}
\caption{The exceptional line (EL) in the $a$-$\mu$ parameter space.
The existence of the EL reflects the fact that an EP is achieved at $k=j=m=0$ without the need for fine-tuning.
}
\label{EL_2D}
\end{figure}

The result shows that the excitation factors diverge at EP, as is expected.
Actually, other spin parameters also admit EPs in the parameter space spanned by $\mu$ and $a$.
We find that EPs form a continuous set (one-dimensional manifold) in the $a$-$\mu$ parameter space, which is an EL (FIG.~\ref{EL_2D}).
No AC parameter is found in our parameter setting, where $a$ and $\mu$ are real and admit the BH solution.

In our system, the emergence of an EL, rather than isolated EPs, is a direct consequence of the $m=0$ symmetry, i.e.,  $\omega_n^{-}=-(\omega_n^{+})^*$ and $A_{kjm}^{-}=(A_{kjm}^{+})^*$. 
This symmetry reduces the EP condition to a single constraint (codimension one~\cite{Heiss:2012dx}), which generically traces out a curve in the two-dimensional $(\mu, a)$ space.
For $m\neq 0$ with $a\neq 0$, this symmetry is lost, and two independent conditions are required (codimension two), yielding at most isolated EPs.
In the spherically symmetric case $a=0$, the symmetry persists even for $m\neq 0$ since $A_{kjm}$ is independent of $\omega$ [Eq.~\eqref{SCd4}]; however, $A_{kjm}>0$ for $m\neq 0$ makes the square root in Eq.~\eqref{anaQNM} strictly positive for any real $\mu$, preventing the EP condition for the same overtones from being satisfied within the physical parameter range.
For example, isolated EPs for $m\neq 0$ in massive scalar perturbations of a Kerr black hole have been found in the $(\ell,m)=(1,1)$~\cite{Cavalcante:2024swt} and $(2,2)$~\cite{Cavalcante:2025abr} sectors.

The existence of EL in the $a$-$\mu$ parameter space is interesting.
It implies that there may be a possibility that we can avoid fine-tuning of BH parameters to observe the excitation of (nearly) double-pole QNMs.
However, when many BH parameters are involved, the relevant parameter space can still be significantly large, and severe fine-tuning may nevertheless remain unavoidable.
Actually, a similar limitation is already inherent in our analysis in the Nariai limit, since it effectively focuses on fine-tuned configurations in which two horizons are extremely close.

Unlike Hermitian degeneracies (Diabolic Points), an EP in non-Hermitian systems is a spectral singularity where not only the eigenvalues (e.g., QNM frequencies) but also their corresponding eigenvectors (e.g., QNM eigenfunction) degenerate.

Heiss has demonstrated in Ref.~\cite{Heiss:2012dx} that in the vicinity of a second-order EP, the eigenvalues $E(\lambda)$ depend on the system parameter $\lambda$ via a square-root expansion (Puiseux series):
\begin{align}\label{GenE}
    E(\lambda) \approx E_{\rm EP} \pm c \sqrt{\lambda - \lambda_{\rm EP}}
\end{align}
This confirms that the spectrum possesses the topological structure, a square-root branch point, in the Riemann surface.
This structure leads to the following phenomenon in the QNM spectrum. 
As discussed in \cite{Cavalcante:2024swt,Cavalcante:2024kmy,Yang:2025dbn,PanossoMacedo:2025xnf}, encircling the EP in the parameter space results in the exchange of the prograde and the retrograde mode, which is caused by the branch point at the EP.
This ``hysteresis'' effect, an important signal of the existence of EP, has been studied in \cite{Cavalcante:2024swt,Cavalcante:2024kmy}.

In our case, the formula of QNM frequencies in Eq.~\eqref{anaQNM} has the square root $\pm\sqrt{V_0/\kappa_{\text{h}}^2-1/4}$, which leads to the branch point [see Eq.~\eqref{GenE}].
It also means that the prograde and retrograde modes (at $k=j=m=0$) in the PT potential share the same EP.

\begin{figure}[t]
\centering
\includegraphics[width=\columnwidth, trim=0.55cm 0.5cm 0.5cm 0.6cm]{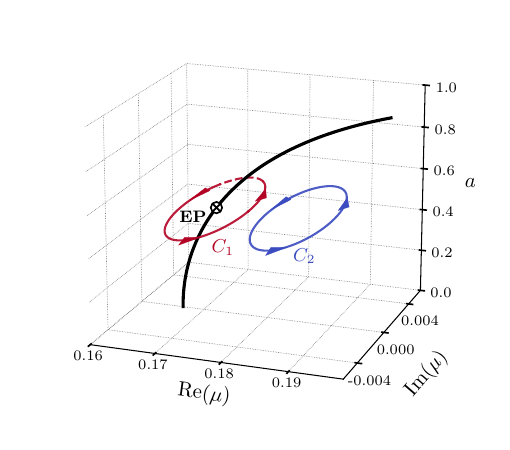}
\caption{EL (black solid) in the parameter space covered by the spin parameter $a$, $\text{Re}(\mu)$, and $\text{Im}(\mu)$.
The parameter trajectories $C_1$ (red) and $C_2$ (blue) restricted on the surface $a=0.5$ are also shown.
The cross marker stands for the EP at $a=0.5$.
}
\label{Param_space}
\end{figure}

FIG. \ref{Param_space} shows the EL in the parameter space (the black line). 
To encircle the EL, we perform an analytic continuation with $\mu \in {\mathbb C}$.
$C_1$ in FIG. \ref{Param_space} shows the parameter trajectory encircling the EL on the plane at $a=0.5$.
$C_2$ is a closed trajectory that does not encircle the EL.
The trajectory of $C_i$ is parameterized as follows:
\begin{align}\label{Param_change}
    (\mu_i + R\cos(2\pi s), R\sin(2\pi s), a), \quad s\in[0,1], 
\end{align}
where $\mu_1 = 0.17211$, $\mu_2 = 0.185$, $R=0.005$, and $a=0.5$.

\begin{figure}[t]
\centering
\includegraphics[width=0.95\columnwidth, trim=0.5cm 0cm 0cm 0cm]{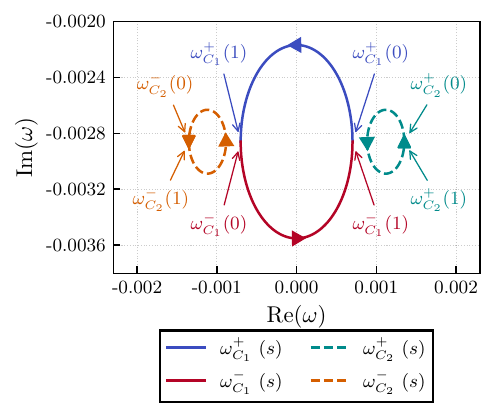}
\caption{Trajectories of QNMs associated with the parameter trajectories $C_1$ (solid) and $C_2$ (dashed).
A closed loop in the parameter space results in an interchange of prograde and retrograde modes if it encircles an EP ($C_1$); otherwise, each QNM returns to its original state after a full cycle ($C_2$).
}
\label{QNM_trajectory}
\end{figure}

\begin{figure}[t]
    \centering
    \includegraphics[width=0.49\linewidth]{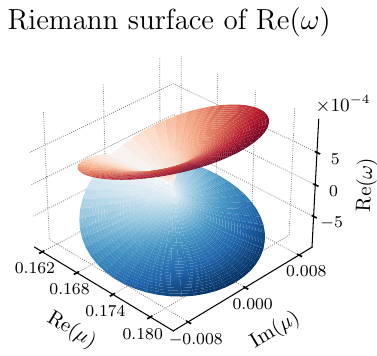}
    \hfill
    \includegraphics[width=0.49\linewidth]{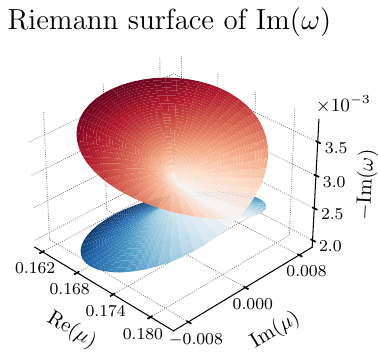}
    \caption{$\sqrt{z}$-type Riemann surface structure of QNMs for $a=0.5$.
    Due to this topology, a complete loop around the EP results in an interchange between the prograde and retrograde modes.
    }
    \label{RiemannStructure}
\end{figure}

FIG.~\ref{QNM_trajectory} shows the trajectories of the fundamental QNMs associated with the parameter trajectory $C_i$. 
QNMs $\omega^{\pm}_{C_i}(s)$ 
are calculated at the parameters on $C_i(s)$ in the range of $0 \leq s \leq 1$, and the superscript of $+ / -$ stands for the prograde/retrograde modes at $s=0$. 
It can be seen that the two QNMs are swapped for $C_1$, which is the hysteresis effect, whereas the hysteresis is not observed for $C_2$. 
This indicates that $C_1$ encircles a branch point, EP, reflecting the $\sqrt{z}$-type Riemann surface structure of the QNM frequencies (FIG.~\ref{RiemannStructure}).

%%%%%%%%%%%%%%%%%%%%%%%%
\section{Conditions for the dominant linear growth of (nearly) double-pole QNMs}
\label{Sec_conditions}
%%%%%%%%%%%%%%%%%%%%%%%%

We here discuss some conditions for the linear growth of QNMs \cite{Yang:2025dbn,PanossoMacedo:2025xnf} near an EP to be dominant in the early stage of ringdown.
To derive \eqref{h_nearEP} from \eqref{h_nearEP_original}, we perform a Taylor expansion of $A_{\rm out}(\omega)$ and $f(\omega)$ around the two modes, whose frequencies differ by $\delta \omega$.
This approximation is valid up to first order provided that $\delta \omega$ is smaller than the characteristic variation scale of $A_{\rm out/in} (\omega)$, which we denote as $\omega_{\rm G}$.
Therefore, we have the first condition:
\begin{align}\label{dw<<wG}
    |\delta\omega| \ll \omega_\text{G},
\end{align}
We would conjecture that in most cases, the scale $\omega_{\rm G}$ would be set by a typical BH scale, such as $(2M)^{-1}$, the surface gravity $\kappa_{\rm h}$, or the {\it typical} frequency separation of discrete QNM spectra.
Indeed, for the Nariai system with $a=0$, \eqref{ain_aout} shows that the frequency dependence of $A_{\rm out / in}$ appears only through the combination $\omega / \kappa_{\rm h}$, i.e., $\omega_{\rm G} = \kappa_{\rm h}$ at least in the Nariai case we considered previously.
On the other hand, the factor $e^{-i \Omega_{\rm b} t}$ in \eqref{h_nearEP_original} should be expanded with $\delta \omega t$.
Then the relevant time scale is found to be
\begin{align}\label{tobs<<1/dw}
    t \ll \text{min} \left[ \frac{1}{|\delta\omega|}, \frac{1}{|\text{Im} (\omega_{\rm EP})|} \right] \,,
\end{align}
where $t \ll 1/|\text{Im} (\omega_{\rm EP})|$ is another condition for the relevant time scale so that the linear growth appears before the mode is exponentially damped.
We here do not regard Eq.~\eqref{tobs<<1/dw} as a condition, since the inequality can always be satisfied by choosing $t$ arbitrarily small.
Nevertheless, in realistic situations such as observations, one cannot take arbitrarily small $t$, because the time resolution must be sufficiently shorter than the relevant timescale.
In this sense, the above inequality \eqref{tobs<<1/dw} still represents a practically important condition.

There should exist another condition for the dominant linear growth of QNM excitation, as the (nearly) double-pole QNM is excited with the following waveform:
\begin{equation}
(iX + Y t) e^{-i \omega_{\rm EP} t}\,,
\label{XY}
\end{equation}
where constants $X$ and $Y$ can be read from our results in \eqref{h_nearEP} or in \eqref{h_EP} (see also Ref.~\cite{PanossoMacedo:2025xnf}).
That is, the second condition under which the linear-growth term, $Yt e^{-i \omega_{\rm EP} t}$, dominates the waveform is
\begin{equation}
\frac{Y}{|\text{Im}(\omega_{\rm EP})|} \gtrsim X  \Rightarrow \frac{1}{|\text{Im}(\omega_{\rm EP})|}\gtrsim \frac{d}{d \omega} \log \left(\frac{A_{\rm out}}{f} \right)\Bigg|_{\omega=\omega_{\rm EP}}\,,
    \label{A_LG_A_DS}
\end{equation}
where we imposed $t \lesssim 1/|\text{Im}(\omega_{\rm EP})|$ and read 
\begin{align}
\begin{split}
X &= - \frac{d}{d\omega}\left( \frac{A_\text{out}}{f}\right)_{\omega=\omega_{\rm EP}}\,,\\ 
Y &= -\left( \frac{A_\text{out}}{f}\right)_{\omega=\omega_{\rm EP}}\,,
\end{split}
\end{align}
from Eq.~\eqref{h_nearEP}.
We further rewrite Eq.~\eqref{A_LG_A_DS} as
\begin{equation}
    q \gtrsim 1\,,
    \label{Amp_ratio}
\end{equation}
where
\begin{equation}
    q \coloneqq \Bigg|\text{Im}(\omega_{\rm EP})\frac{d}{d \omega} \log \left(\frac{A_{\rm out}}{f} \right)\Bigg|^{-1}_{\omega=\omega_{\rm EP}}\,.
    \label{q_definition}
\end{equation}
The larger $q$ is, the more dominant the linear growth term is, provided that the condition in Eq.~\eqref{dw<<wG} is sufficiently satisfied.
Under the two conditions, Eqs.~\eqref{dw<<wG} and \eqref{Amp_ratio}, the linear growth of nearly double-pole QNMs, $\sim te^{-i\omega t}$, can be dominant in the early ringdown.
The above discussions and the derived conditions apply to a broader class of systems.
However, we emphasize that the above conditions take into account only the two QNMs that can become degenerate, and do not include the excitation of other QNMs.
We now examine the linear growth of the QNM excitation in our specific system, involving the Nariai BH and a massive scalar field, and verify whether those conditions are satisfied.
Setting $a=0$ for simplicity in our model, the conditions (\ref{dw<<wG}) becomes:
\begin{align}\label{dw<<wG2}
    |\delta\omega| \ll \omega_\text{G} = \kappa_\text{h}\,.
\end{align}
To verify this condition, we consider a waveform in Eq. (\ref{hE}) with $N_+=N_-=0$, that is, only with the fundamental prograde and retrograde modes.

\begin{figure}[t]
\centering
\includegraphics[width=\columnwidth]{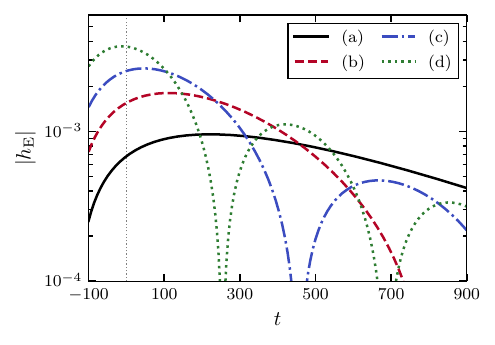}
\caption{The fundamental prograde and retrograde modes whose interference causes the linear growth in the early stage of ringdown. The time-domain waveforms are plotted for the cases of (a)-(d) in Table~\ref{tab:parameters}.
The vertical gray dashed line stands for the starting time of the ringdown $t=0$.
}
\label{Ringdown_Comparison}
\end{figure}

FIG.~\ref{Ringdown_Comparison} shows the waveforms for the four values of $\mu$ listed in Table.~\ref{tab:parameters}. 
In the case of (a), $|\delta\omega| / \omega_\text{G} = 3.41 \times 10^{-3}$ is much smaller than unity and $q$ is larger than unity (see Table~\ref{tab:parameters}), as such the both conditions in Eqs.~\eqref{dw<<wG} and \eqref{Amp_ratio} are satisfied.
Indeed, in this case, the linear growth dominates the signal around $t \sim 0$ (black solid in FIG.~\ref{Ringdown_Comparison}).
As the value of $|\delta \omega| / \omega_{\rm G}$ increases and exceeds unity---namely (b) $|\delta \omega| / \omega_{\rm G} = 1.18$, (c) $1.82$ and (d) $2.62$---the linear growth becomes less dominant (FIG.~\ref{Ringdown_Comparison}).

\begin{figure}[t]
\centering
\includegraphics[width=\columnwidth]{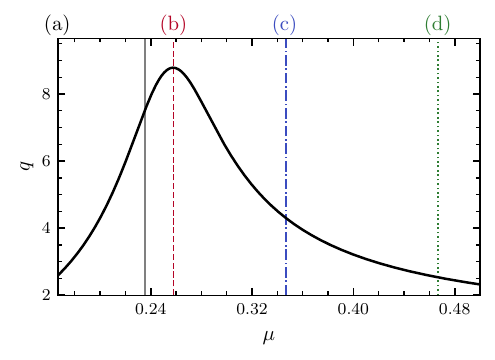}
\caption{Plot of $q$ in Eq.~\eqref{q_definition} with respect to $\mu$. The gray line represents the value of $\mu$ at which $|\delta\omega| \simeq \omega_\text{G}$.
Case (a) has $\mu = 1/6 + 10^{-10}$, which lies at the left edge of the panel.
The maximum value of $q$ occurs in case (b), with $\mu \simeq 0.258$ (red dashed), where the second condition $q \gtrsim 1$ \eqref{Amp_ratio} is strongly satisfied.
The blue dot-dashed and green dotted lines correspond to cases (c) and (d), respectively.
}
\label{q_vs_mu}
\end{figure}

For the cases we examined, the ratio $q$ is of the order of unity and $q > 1$, which indicates that the linear growth, $Yt e^{-i \omega_{\rm EP} t}$, dominates the damped oscillation $X e^{-i \omega_{\rm EP} t}$ in Eq.~\eqref{XY}.
Interestingly, we find a nontrivial dependence of $q$ on the scalar mass $\mu$ (FIG.~\ref{q_vs_mu}).
The maximum value of $q$, at which the second condition is strongly satisfied, appears slightly beyond $\mu \simeq 0.2357$, where the first condition is violated, i.e., $|\delta\omega| = \omega_{\rm G}$, (gray solid line in FIG.~\ref{q_vs_mu}). 

We propose that the ratio $q$ in Eq.~\eqref{Amp_ratio} serves as a proper indicator to confirm the dominance of the linear growth, provided that condition \eqref{dw<<wG} is satisfied. 
It would be interesting to evaluate the ratio $q$ as well as $|\delta \omega| / \omega_{\rm EP}$ for other BH systems which exhibit EPs or ELs.

\begin{table}[t]
    \centering
    \caption{Parameters relevant to the conditions Eqs.~\eqref{dw<<wG} and \eqref{Amp_ratio}. 
    With our parameter setting, $\omega_\text{G}=\kappa_\text{h}=5.56 \times 10^{-3}$.
    }
    \label{tab:parameters}
    \begin{tabular}{lccc} %
        \midrule
        \midrule
        label & $\mu$ & $|\delta \omega| / \omega_{\rm G}$ & $q$\\
        \midrule
        (a) & $1/6 + 10^{-6}$ & $3.41\times10^{-3}$ & $2.59$ \\
        (b) & $0.258$ & $1.18$ & $8.79$\\
        (c) & $1/6+9/50$ & $1.82$ & $4.30$ \\
        (d) & $1/6+3/10$ & $2.62$ & $2.53$ \\
        \midrule
        \midrule
    \end{tabular}
\end{table}

%%%%%%%%%%%%%%%%%%%%%%%%
\section{Conclusion}
%%%%%%%%%%%%%%%%%%%%%%%%
\label{sec_conclusions}

We have considered the Nariai limit in the Myers–Perry–de Sitter black hole (BH) with a single rotation to analytically investigate quasinormal mode (QNM) excitation near and on exceptional points (EPs) and exceptional lines (ELs).
In doing so, we have identified a phenomenon in which parameters such as the BH spin and the scalar mass form an EL.
We have also carefully demonstrated that a wave-scattering in the P\"{o}schl–Teller (PT) potential provides a useful framework for analytically analyzing EPs or ELs.

By analyzing the radial perturbation equation, which reduces to a wave equation with the PT potential, we have analytically constructed the Green’s function and computed the excitation factors (see also Ref.~\cite{Berti:2006wq}).
Then, we have clarified how EPs/ELs arise in our setup.
We have shown that, regardless of the spacetime dimension, the prograde and retrograde modes with the same overtone number can degenerate, leading to the divergence of their excitation factors.
Despite the divergence, the time-domain waveform remains stable against the parameter variations near the EP, consistent with the destructive excitation of near double-pole QNMs \cite{Oshita:2025ibu}. 
The amplitude stays finite and does not signal any instability of the Nariai spacetime. 
The destructive interference between the prograde and retrograde branches ensures the stability of the reconstructed signal, while transient linear growth in the time domain \cite{Yang:2025dbn} emerges as a unique feature of (nearly) double-pole QNMs.

We have also found that double-pole QNMs exist even for non-zero single rotation.
That is, we have found that EPs form a continuous set, i.e., an EL, in the parameter space spanned by the spin parameter and the scalar-field mass.
Such a distribution of EPs has been discussed in recent work based on the small-bump correction in the Regge-Wheeler potential \cite{Cao:2025afs}.
In general, observing the excitation of (nearly) double-pole QNMs requires significant fine-tuning of BH parameters.
However, in systems that possess an EL in parameter space, QNM degeneracy could occur with reduced fine-tuning.
Note that when many BH parameters are involved, the relevant parameter space can still be considerably large, and severe fine-tuning may nevertheless remain unavoidable in realistic situations, even in the presence of an EL.
Actually, a similar limitation is already inherent in our analysis in the Nariai limit, since it effectively focuses on configurations in which two horizons are extremely close, which itself corresponds to a finely tuned region of the parameter space.
For this reason, the advantage of having an EL becomes more significant when the relevant parameter space is effectively low-dimensional.

On the other hand, the Kerr BH with realistic spin parameters does not possess an EP in $\ell=m=2$ mode, which is the dominant sector in gravitational waves sourced by binary systems.
Instead, only avoided crossings (ACs) occur for overtones of $n =5$ and $n=6$ \cite{Motohashi:2024fwt}, but their decay rates are very large.
Therefore, the observation of the effect of EPs, i.e., linear growth, would be challenging unless the BH geometry is significantly modified by physics beyond general relativity, by strong environmental effects, or by the change in the global structure of the surrounding environment, e.g., a small-bump correction in the far zone.

Our analytic treatment is based on the expansion in $h := r_c - r_h$ and holds as long as $h \ll 1$; away from this regime, the specific EP parameters may shift, and the closed-form expressions cease to hold. 
The symmetry at $m=0$ that reduces the codimension to one follows from the general structure of Eqs.~(\ref{Req1}) and (\ref{Seq1}), not from the P\"oschl-Teller reduction.
This suggests that, at least from a simple counting of the number of constraints and the dimensionality of the parameter space, the EP locus for $k=j=m=0$ forms a codimension-one surface in the full $(\mu M, a/M, \Lambda M^2)$ parameter space. If this is the case, the EL found here can be regarded as a cross-section between the high-dimensional EP locus and the Nariai constraint.

We have carefully studied some necessary conditions under which the transient linear growth, caused by the excitation of (near) double-pole QNMs, to be observed.
The two conditions we derived are particularly novel and important [Eqs.~\eqref{dw<<wG} and \eqref{Amp_ratio}].
At an EP, the QNM excitation amplitude generally appears as a linear combination of a damped oscillation and a (transient) linear-growth term (see Eq.~\eqref{h_EP} and see also Ref.~\cite{PanossoMacedo:2025xnf}).
The conditions we derive provide a general criterion under which the linear-growth term becomes dominant at earlier times.
These conditions are general and apply to a broad class of models in which EPs arise.

The present results provide deeper insight into the spectral behavior of BHs and offer a concrete example in which unique non-Hermitian phenomena arise in gravitational systems. 
Future work may extend this analysis to gravitational perturbations, investigate nonlinear effects near EPs, and explore possible signatures in gravitational-wave observations.
It is also important to assess the detectability of (near) double-pole QNM excitation in ringdowns.

%%%%%%%%%%%%%%%%%%%%%%%%
\section{Data availability}
%%%%%%%%%%%%%%%%%%%%%%%%
The data that support the findings of this article are openly available \cite{Nakamoto2026Data}.

\acknowledgments
N.~N.~was supported by the Create the Future Project at Kyoto University.
N.~O.~was supported by Japan Society for the Promotion of Science (JSPS) KAKENHI Grant No.~JP23K13111 and by the Hakubi project at Kyoto University.

\appendix

%%%%%%%%%%%%%%%%%%%%%%%%
\section{Application to the high-dimensional spinning black hole ($d=5$)}
\label{App_d5_case}
%%%%%%%%%%%%%%%%%%%%%%%%
\begin{figure*}[t]
  \centering
  \includegraphics[width=\linewidth, trim=0cm 0.3cm 0cm 0cm]{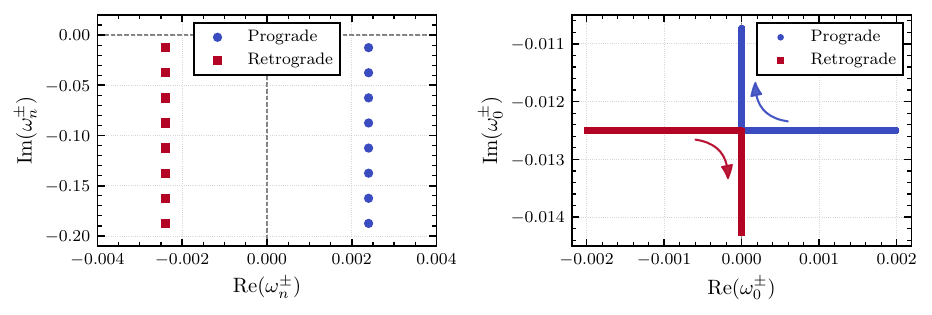}
  \caption{
    ({\bf Left}) Distribution of QNMs for prograde and retrograde modes when $a=0$, $\mu=0.36$, and $k=j=m=0$ up to $n=7$ modes.
    ({\bf Right}) Trajectory of the fundamental QNMs, $\omega_0^{\pm}$, near or at the exceptional point. 
    The value $\mu$ decreases from $0.358$ to $0.35$ along the arrows, and the two QNMs intersect at $\mu=1/2\sqrt{2}$.
  }
  \label{fig:qnm_combined5}
\end{figure*}
In this Appendix, we demonstrate that our analysis of the EL in the main text can be extended to a high-dimensional spacetime.
As an example, we here analyze the QNM excitation near or at EP in $d=5$.
In the Nariai limit with $d=5$, the spin parameter $a$ and $\Lambda$ has the relation:
\begin{align}
    \Lambda(a) = \frac{1}{-a^2+4+2\sqrt{-2a^2+4}}\,.
\end{align}
The horizon radius in the Nariai limit $r_{\rm h} (= r_{\rm c})$ can then be expressed as
\begin{align}
    r_{\rm h}=\sqrt{(1-a^2\Lambda(a))/2\Lambda(a)}.
\end{align}

In the case of $a=0$, there exists an EP, at which all prograde and retrograde QNMs degenerate. 
Denoting the difference between the outer horizon $r_{\text{h}}$ and the cosmological horizon $r_{\rm c}$ as $h \coloneqq r_{\text{c}}-r_{\text{h}}$, the massive scalar perturbation equation can be written as Eq.~\eqref{V_0_formula}, where $\kappa_{\rm h}$ and $V_0$ are
\begin{align}
    \kappa_{\text{h}} &\simeq \frac{h}{4}, \\
    V_0 &\simeq \frac{h^2}{32}(4\mu^2+A_{kjm})\,.
\end{align}
Eq.~\eqref{V_0_formula} describes wave scattering with the PT potential, whose Green's function has the analytic form.
The QNM frequencies and the eigenvalue of the separation constant are 
\begin{align}
    \omega_n^{\pm} &\simeq \pm\frac{h}{4}\sqrt{\frac{1}{2}(4\mu^2+A_{kjm})-\frac{1}{4}}-\frac{hi}{4}(n+\frac{1}{2})\,, \label{PTQNMd5}\\
    A_{kjm}&=(2k+j+m)(2k+j+m+2)\,. \label{SCd5}
\end{align}
Note that the parameter $h$ serves as the expansion parameter of the Nariai limit, so we take the value of $h=0.1$. 
In this case, QNMs are distributed with equal intervals along the imaginary axis in the complex frequency plane (left panel in FIG.~\ref{fig:qnm_combined5}).

For the monopole mode $k=j=m=0$, we can make the term under the square root in (\ref{PTQNMd5}) zero by tuning the other degree of freedom $\mu$.
Then the degeneracy between the prograde and retrograde modes arises, which corresponds to the EP (right panel in FIG.~\ref{fig:qnm_combined5}). 
When the spin parameter $a=0$, the separation constant \eqref{SCd5} does not depend on the overtone number $n$, so the prograde and retrograde modes degenerate at the same mass $\mu=1/2\sqrt{2}$ for all overtone numbers.

\begin{figure}[!t]
\centering
\includegraphics[width=\columnwidth]{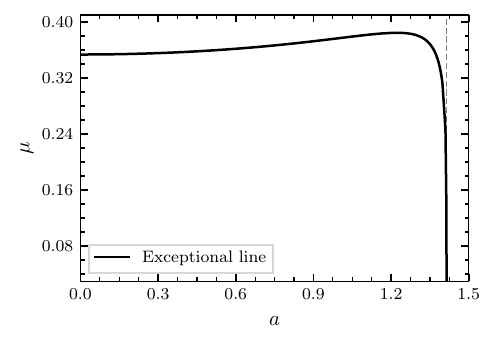}
\caption{The exceptional line (EL) in the $a$-$\mu$ parameter space.
The gray dashed line stands for the upper bound in Eq.~\eqref{upper_spin_d5}.
}
\label{EL_2D_d=5}
\end{figure}

We confirm that even in the high-dimensional ($d=5$) and spinning BH, $a>0$, EP still arises for the monopole mode $k=j=m=0$.
It can be numerically confirmed by searching for the parameters leading to $V_0/\kappa_{\rm h}^2 = 1/4$ at the eigenvalues $\omega_n$ and $A_n$, where
\begin{align}
    \frac{V_0}{\kappa_{\text{h}}^2} &= \frac{1}{r^2_{\text{h}}(d-1)\Lambda} \notag \\
    &\times \left(\frac{j(j+d-5)a^2}{r^2_{\text{h}}} + \mu^2 r^2_{\text{h}} + A_{n}\right)\,.
\end{align}
That is, we find that the EPs form an EL in the parameter space spanned by the spin parameter $a$ and the scalar mass $\mu$ (FIG.~\ref{EL_2D_d=5}).

%%%%%%%%%%%%%%%%%%%%%%%%
\section{The divergence of excitation factors}
\label{Sec_div_EFs}
%%%%%%%%%%%%%%%%%%%%%%%%

We here discuss the divergence of excitation factors for the Nariai limit in an analytic way.
The perturbation we consider is restricted to $k=j=m=0$.
To this end, let us consider the degeneracy between the prograde and retrograde QNMs with the same overtone number, whose frequencies are $\omega_n^{+}$ and $\omega_n^{-}$ respectively. 
The conditions for these QNM frequencies and the eigenvalues of the separation constant to degenerate are [see Eq.~\eqref{anaQNM}]
\begin{align}
    &\omega_n^{+} = \omega_n^- \Leftrightarrow  \sqrt{\frac{V_0^{+}}{\kappa_{\text{h}}^2}-\frac{1}{4}} + \sqrt{\frac{V_0^{-}}{\kappa_{\text{h}}^2}-\frac{1}{4}} = 0, \label{exponent_cond1} \\
    \label{eq:sum_sqrt}
    &A_{n}^{+} = A_{n}^{-}\,.
\end{align}
We find that $A_{n}^{+/-}$ is real-valued from the symmetry relation $A_{n}^{-}=(A_{n}^{+})^*$.
Remember that we consider $k=j=m=0$.
Then, \eqref{eq:sum_sqrt} makes the terms under the square root in \eqref{exponent_cond1} identical [see Eq.~\eqref{V_0_formula}]:
\begin{align}
    \frac{V_0^{+}}{\kappa_{\text{h}}^2}-\frac{1}{4} = \frac{V_0^{-}}{\kappa_{\text{h}}^2}-\frac{1}{4}\,.
    \label{identical_squareroot_exponent}
\end{align}
From Eqs.~\eqref{exponent_cond1} and \eqref{identical_squareroot_exponent}, we find [see Eq.~\eqref{beta_definition}]
\begin{align}
\begin{split}
    &\sqrt{\frac{V_0^{+}}{\kappa_{\text{h}}^2}-\frac{1}{4}} = \sqrt{\frac{V_0^{-}}{\kappa_{\text{h}}^2}-\frac{1}{4}} = 0, \label{sqrtzero} \\
    &\Leftrightarrow \beta = -\frac{1}{2}.
\end{split}
\end{align}
Plunging $\beta = -1/2$ into the formula of the excitation factors in Eq.~\eqref{PT_excitation_fac_beta}, we find that the factors $\Gamma(-n-2\beta-1)$ and $\Gamma(1+2\beta-n)$ in Eq.~\eqref{PT_excitation_fac_beta} diverge, i.e., the divergence of $B_n^{\pm}$ at the degeneracy of QNM frequencies.
We also numerically verified the divergent behavior of excitation factors in FIGs.~\ref{fig:4plots1} and \ref{QNM_EF_a>0}.

\bibliography{reference}

\end{document}